\documentclass[sigconf,nonacm,balance=false]{acmart}
\setcopyright{none}
\AtBeginDocument{}
\usepackage{tabularx}
\usepackage{booktabs}
\usepackage{multirow}
\renewcommand{\dbltopfraction}{0.85}
\renewcommand{\dblfloatpagefraction}{0.65}
\renewcommand{\textfraction}{0.1}
\allowdisplaybreaks[1]
\usepackage{fvextra}
\DefineVerbatimEnvironment{PromptBlock}{Verbatim}{
  fontsize=\fontsize{7}{8.4}\selectfont,
  fontfamily=tt,
  breaklines=true,
  breakanywhere=true,
  breaksymbolleft={},
  breakanywheresymbolpre={},
  frame=none,
  tabsize=2
}
\begin{document}
\title{EgoAsk: Egocentric Teaching of Personalized Object Knowledge for Household Robots}

\author{Yuanda Hu}
\affiliation{
  \department{College of Design and Innovation}
  \institution{Tongji University}
  \city{Shanghai}
  \country{China}
}
\email{ydhu@tongji.edu.cn}

\author{Wenbin Zuo}
\affiliation{
  \department{Shanghai Research Institute for Intelligent Autonomous Systems}
  \institution{Tongji University}
  \city{Shanghai}
  \country{China}
}
\email{wbzuo@tongji.edu.cn}

\author{Yiting Shen}
\affiliation{
  \department{College of Design and Innovation}
  \institution{Tongji University}
  \country{China}
}
\email{2350215@tongji.edu.cn}

\author{Tianle Chen}
\affiliation{
  \department{School of Computer Science and Technology}
  \institution{Tongji University}
  \city{Shanghai}
  \country{China}
}
\email{2634280@tongji.edu.cn}

\author{Hector Fabio Calero Tobar}
\affiliation{
  \department{College of Design and Innovation}
  \institution{Tongji University}
  \city{Shanghai}
  \country{China}
}

\author{Yate Ge}
\affiliation{
  \department{College of Design and Innovation}
  \institution{Tongji University}
  \city{Shanghai}
  \country{China}
}
\email{geyate@tongji.edu.cn}

\author{Xiaohua Sun}
\affiliation{
  \department{School of Design}
  \institution{Southern University of Science and Technology}
  \city{Shenzhen}
  \state{Guangdong}
  \country{China}
}

\author{Weiwei Guo}
\affiliation{
  \department{College of Design and Innovation}
  \institution{Tongji University}
  \city{Shanghai}
  \country{China}
}
\email{weiweiguo@tongji.edu.cn}

\renewcommand{\shortauthors}{Hu et al.}

\begin{abstract}
Unlike users, who know their own belongings and routines, household robots cannot easily acquire such personalized object knowledge automatically and depend on users to teach them. User-initiated teaching requires users to arrange dedicated teaching sessions and decide what to teach, even when they are unsure what the robot needs to learn. We introduce \textit{EgoAsk}, a smart-glasses-based system that proactively embeds personalized object teaching into everyday activities. EgoAsk shares the user's first-person view with the robot, identifies gaps in personalized object knowledge, and analyzes ongoing activity to ask context-relevant questions that support future household assistance. To examine how teaching initiative and question timing affect users' teaching experiences, we conducted a within-subjects study with 18 participants and found lower reported knowledge-gap monitoring burden with robot-initiated questioning and less need for context reconstruction with EgoAsk. These findings characterize teaching burdens and timing preferences, offering design implications for egocentric robot-teaching systems.
\end{abstract}



\maketitle
\raggedbottom

\begin{figure*}[t]
  \includegraphics[width=\textwidth]{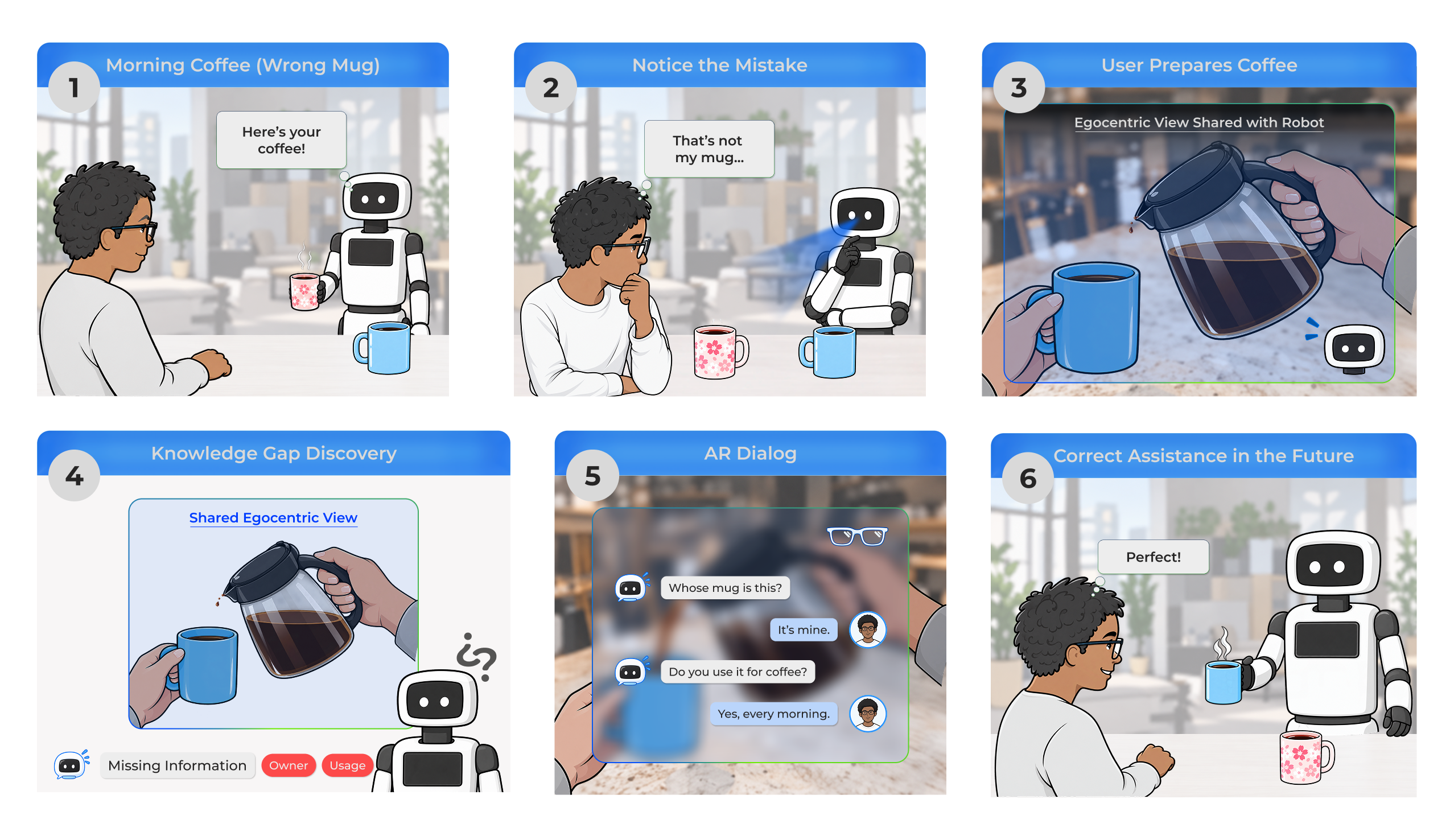}
  \caption{A morning with EgoAsk: (1--2) At the dining table, a robot serves coffee in the wrong mug because it does not know which mug belongs to the user or which one they usually use for coffee. (3) On a later occasion, the user prepares coffee in the kitchen while EgoAsk shares their first-person view with the robot through smart glasses. (4--5) Using this shared view, the robot identifies gaps in its knowledge about the mug and proactively asks context-relevant questions. The user explains that the mug is theirs and that they use it for coffee. (6) This knowledge can support future personalized assistance, such as selecting the user's own mug for morning coffee. EgoAsk is a smart-glasses-based system that enables users to teach household robots personalized object knowledge through proactive, context-relevant questions during everyday activities.}
  \label{fig:teaser}
\end{figure*}

\section{Introduction}
\label{sec:intro}

Users accumulate personalized knowledge about specific objects in everyday life, including ownership, usage habits, and placement preferences~\cite{kwon2026memento}. Household service robots need to learn this knowledge to provide assistance that aligns with users' needs and routines~\cite{wu2023tidybot,dai2024think}. However, knowledge that is familiar or even self-evident to users may remain unknown to a robot. For example, when asked to ``put my cup back,'' a robot may recognize the cups on the table but still not know which one to pick up or where to place it. Current vision-language models can recognize objects and understand their general semantics, but they cannot reliably infer such person-specific knowledge from appearance alone~\cite{dai2024think,hashimoto2026actowl}. Even with examples previously provided by the user, a robot may incorrectly generalize the user's preferences~\cite{wu2023tidybot}. Robots therefore need to acquire personalized object knowledge through interaction with users to support subsequent household tasks~\cite{dai2024think,hashimoto2026actowl}.

Early interactive object-teaching systems required users to present objects within the robot's field of view and provide labels, limiting teaching to the robot's location and visual coverage~\cite{pasquale2015teaching}. Smart glasses expanded teaching beyond the robot's field of view, enabling users to indicate objects in the environment through gaze or inspect and correct recognition results through a mixed-reality interface~\cite{weber2023multiperspective,jaykumar2024iteach}. Although these interfaces make object reference and feedback more convenient, user-initiated teaching workflows still present two challenges. (1) Even when teaching can take place in situ, users must select an object and initiate an exchange, organizing an additional teaching activity to support the robot's learning~\cite{pasquale2015teaching,ayub2024interactive,jaykumar2024iteach}. (2) Users must decide what to teach without being able to align their teaching with the robot's actual knowledge gaps~\cite{cakmak2010designing}. 
These challenges motivate our design goal: how to enable users to teach robots personalized object knowledge during everyday activities without needing to organize dedicated exchanges or decide what to teach.

To pursue this goal, we draw on research in interactive object learning, robot active learning, and egocentric perception and interaction to derive three design considerations. First, robots should share the user's first-person view, reducing users' need to accommodate the robot's location and field of view~\cite{cai2025aiget}. Second, robots should proactively ask about personalized knowledge gaps relevant to future household assistance, clarifying what information users need to provide~\cite{chao2010transparent,hashimoto2026actowl,liang2026pahf}. Third, questions should relate to how users are using or handling objects, connecting knowledge exchange with everyday object interactions~\cite{cai2025aiget,li2025satori}.

Guided by these considerations, we designed and implemented \textit{EgoAsk}, a system that uses smart glasses to help household robots learn personalized object knowledge. EgoAsk supports teaching during everyday activities, allowing users to discuss nearby objects with the robot without separately presenting them to it. By sharing the user's first-person view, EgoAsk enables the robot to identify the specific objects the user interacts with, understand the associated activities, and proactively ask about personalized knowledge gaps. Users can teach by answering questions or volunteering additional knowledge. For example, while a user pours coffee, EgoAsk can ask about the cup's ownership and everyday use. An answer such as ``This is my cup; I use it for coffee every morning'' provides knowledge that can help the robot later select a cup consistent with the user's habits (Figure~\ref{fig:teaser}).

To determine what personalized object knowledge to learn from the user, EgoAsk first identifies a target object in the first-person view and marks it on the smart glasses' heads-up display (HUD). It then retrieves household assistance tasks involving that object within the robot's capabilities, identifies personalized knowledge needs that could affect future assistance, and compares these needs with existing knowledge to identify candidate knowledge gaps. In parallel, EgoAsk analyzes successive frames to understand the user's ongoing activity involving the target object. Using this activity context, it selects candidate knowledge gaps that are relevant to the current activity and useful for the robot's future assistance~\cite{cai2025aiget,li2025satori}, then generates specific questions. Each question addresses a single knowledge need and is presented through speech, with users responding verbally. EgoAsk stores their answers as personalized object knowledge, enabling the robot to retrieve suitable objects in later tasks~\cite{liang2026pahf}.

We conducted a within-subjects study with 18 participants in a simulated home environment to compare three teaching approaches in an object-tidying task: (1) User-Led Teaching, where users initiated teaching; (2) Post-Task Questioning, where the robot asked questions after the task; and (3) In-Task Questioning (EgoAsk), where the robot asked questions during the task. Overall workload did not differ significantly across the three conditions, while specific teaching burdens varied. Compared with User-Led Teaching, both robot-initiated conditions reduced knowledge-gap monitoring burden, helping participants identify what remained to teach. Participants reported less need to reconstruct context with EgoAsk than with Post-Task Questioning, as interviews indicated they could draw on visible objects and ongoing actions when answering. Although some participants reported that in-task questions interrupted their actions, 11 of the 18 participants (61\%) ranked EgoAsk first. Interviews also revealed individual differences in timing preferences, with some participants favoring answering in context and others preferring to defer until the task was complete. These findings suggest that integrating personalized object teaching into everyday activities requires attention to the robot's knowledge needs, the contextual cues available when users answer, and users' choice of when to engage.

Our contributions are threefold:
\begin{itemize}
  \item Three design considerations that enable users to teach robots personalized object knowledge during everyday activities, without organizing dedicated sessions or deciding what to teach.
  \item EgoAsk, a smart-glasses-based system that turns everyday object interactions into teaching opportunities for household robots and retains the acquired knowledge for future assistance.
  \item Empirical findings on how teaching initiative and question timing redistribute teaching burdens across knowledge-gap monitoring, context reconstruction, and activity interruption, together with design implications for future robot teaching systems.
\end{itemize}

\section{Related Work}
\label{sec:related-work}
\subsection{Personalized Object Knowledge for Household Robots}
\label{sec:rw-personalization}

Unlike general object properties, personalized object knowledge is specific to individual users~\cite{kwon2026memento}. It includes object ownership and the associated social norms governing how robots should interact with people's belongings~\cite{tan2019mine,hu2023interactive}, as well as the emotional value and user-specific functional roles assigned to particular objects~\cite{yin2026hold}. Personalization also encompasses preferences for object placement and arrangement~\cite{kapelyukh2022myhouse,wu2023tidybot}, together with patterns of object use within everyday routines~\cite{kwon2026memento}.

Robots can acquire this knowledge by observing users and their interactions with objects. Ownership-learning methods infer relations between users and their belongings by identifying individuals and the object instances they carry~\cite{wu2020ownership}. Spatial preference learning uses observed object arrangements, including historical configurations and the current scene, to infer how users organize their belongings~\cite{kapelyukh2022myhouse,ramachandruni2025parsec}. Related methods infer placement rules from a small number of user-provided examples and generalize them to new objects~\cite{wu2023tidybot}. Temporal models extend these observations to sequences of object locations, learning movement patterns associated with daily routines~\cite{patel2023spatiotemporal}. Combining object histories with user actions further enables predictions of subsequent object use~\cite{patel2023routine}.

Robots can then use this knowledge to adapt household assistance to individual users' needs and preferences. Ownership information constrains which actions robots may perform on users' belongings~\cite{tan2019mine}, while user-specific descriptions and interaction feedback support personalized object search~\cite{dai2024think}. Learned spatial preferences guide object placement in tidying and table-setting tasks~\cite{wu2023tidybot,newman2024degustabot}. Carrying out these tasks also requires grounding user-specific requests in the intended object instances. Related work uses visual and textual references for instance navigation~\cite{barsellotti2024personalized}, and visual references to distinguish and manipulate user-specified objects among similar alternatives~\cite{lee2026bringmycup}.

\subsection{User-Initiated Teaching and Robot-Initiated Questioning}
\label{sec:rw-teaching}

Household robots have difficulty acquiring personalized object knowledge autonomously~\cite{wu2023tidybot,dai2024think} and therefore rely on users to teach them. Prior work on interactive object learning has enabled users to teach robots by presenting objects and providing their names~\cite{fanello2013icubworld,pasquale2015teaching}. Continual object learning extends this interaction over repeated encounters, allowing robots to learn additional objects and associated knowledge incrementally~\cite{ayub2020tellme,ayub2024interactive}. Studies of these interactions document how users adopt different strategies for teaching, correcting, and reteaching robots~\cite{ayub2024human}. In these settings, users not only supply knowledge but also decide when to teach and which information to provide or revisit.

Robots can also take initiative in learning by selecting examples for users to label or asking questions to obtain needed information~\cite{cakmak2010designing,cakmak2012designing,maiettini2022handheld}. Beyond acquiring labels, questions can elicit user preferences through comparisons of behavioral alternatives or direct inquiries about desired robot behavior~\cite{sadigh2017active,wang2024apricot}. In visually grounded tasks, questions help clarify ambiguous instructions by identifying the intended object or grasping goal~\cite{dogan2022asking,kang2023prograsp}. Questioning has also been used to acquire personalized object knowledge, with ownership queries selected according to expected information gain and users' answers used to update ownership estimates~\cite{hashimoto2026actowl}. Clarification before an action can further be combined with feedback afterward to update user-specific preference memory over time~\cite{liang2026pahf}.

The learning benefits of robot-initiated questioning do not necessarily translate into lower teaching effort. A comparison of active and passive teaching found that active questioning improved concept learning without significantly reducing teaching time or improving efficiency; the study also highlighted the importance of user control~\cite{cakmak2010designing}. Other work has evaluated query order and semantic relatedness in terms of both learning performance and teachers' response accuracy, response time, and subjective workload~\cite{racca2019teacheraware}. Research on socially situated learning also considers human engagement and responsiveness alongside knowledge acquisition~\cite{krishna2022socially}. We examine how user-initiated teaching and robot-initiated questioning shape the effort involved in conveying personalized object knowledge.

\subsection{Egocentric and AR-Mediated Robot Teaching}
\label{sec:rw-egocentric}

Augmented reality (AR) and mixed reality (MR) interfaces support robot teaching by presenting the robot's perceptions, goals, and explanations within the physical environment~\cite{suzuki2022augmented}. For object categorization, MR interfaces communicate learning outcomes through feedback formats evaluated for usability and subjective workload~\cite{nakamura2022multimodal}. Shared gaze and interactive annotation allow users to indicate target objects and provide supervision for robot perception~\cite{weber2023multiperspective,weber2023saliency,jaykumar2024iteach}. Bidirectional clarification further supports teaching by displaying candidate objects to resolve ambiguous references~\cite{rosen2020bidirectional} or combining user demonstrations with robot questions~\cite{belardinelli2025train}.

Egocentric sensing captures nearby objects and the user's interactions with them from a first-person perspective. Such observations have been used as demonstrations for learning robot manipulation policies~\cite{kareer2025egomimic,liu2025egozero} and for incorporating a user's task execution and spatial viewpoint into robotic assistance~\cite{wang2020see}. Complementary datasets and benchmarks support activity-understanding research through temporal action annotations~\cite{damen2022rescaling} and tasks concerning hand--object interactions and object state changes~\cite{grauman2022ego4d}. Video-language pretraining aligns egocentric video clips with textual descriptions through contrastive learning~\cite{lin2022egovlp}. Generated video narrations can further provide supervision for learning video-language representations~\cite{zhao2023lavila}. Egocentric assistants also use descriptions of audiovisual clips as retrievable context for subsequent question answering~\cite{yang2025egolife}.

Building on these sensing capabilities, wearable assistants use gaze, environmental context, and user profiles to select relevant information~\cite{cai2025aiget}, and combine environmental observations with user-state modeling to offer proactive guidance~\cite{li2025satori}. Context relevance alone, however, does not determine whether a question is well timed. Interruption research links the effects of an interruption to when it occurs within a task~\cite{adamczyk2004not} and uses task structure to predict the cost of resuming interrupted work~\cite{iqbal2006leveraging}. Robot systems have used social and contextual cues to estimate interruptibility and decide when to request assistance~\cite{banerjee2018interruptibility}, while wearable assistance has explored working-memory models to guide proactive prompting~\cite{pu2025promemassist}. These studies highlight the relevance of activity context to both the content and timing of assistance. EgoAsk draws on this perspective to initiate context-relevant questions about user-specific object knowledge during everyday activities.

\section{Design Considerations}
\label{sec:design-considerations}

Our design goal is to integrate the teaching of personalized object knowledge into everyday activities, reducing the need for users to organize separate teaching sessions or decide what to teach. Users' everyday object interactions provide natural opportunities for such exchanges. During these interactions, the system can share the user's first-person view, identify personalized knowledge worth learning, and relate questions to the ongoing activity. Drawing on this goal and prior work, we propose three design considerations.

\subsection*{D1: Enable the robot to share the user's first-person view.}
When teaching depends on the robot's own field of view, users may need to reposition themselves or deliberately present an object so that the robot can see it~\cite{pasquale2015teaching}. Shared gaze and augmented reality have enabled users to indicate objects in the environment and teach object categories to robots~\cite{weber2023multiperspective}, while mixed-reality interfaces have supported inspecting and correcting the robot's perception results~\cite{jaykumar2024iteach}. First-person video captures objects within the user's immediate environment and the actions performed on them~\cite{cai2025aiget}. Sharing this view during everyday teaching can give the robot contextual cues about how the user handles objects. For example, when a user puts away a cup, the shared view can show both the cup and the cabinet it is placed in. The user can then explain, ``This cup usually goes here,'' without bringing the cup over to the robot for a separate demonstration. The system should therefore let the robot share the user's first-person view, reducing the need for users to move or present objects specifically for teaching.

\subsection*{D2: Proactively ask about gaps in personalized knowledge.}
We define personalized object knowledge as user-specific facts and preferences about individual objects, such as their ownership, usual uses, and usual storage locations~\cite{kwon2026memento}. Drawing on research on transparency in robot learning~\cite{chao2010transparent,huang2020nonverbal}, we identify four demands of user-initiated teaching in our setting: (1) determining which personalized information would be useful for subsequent robot assistance; (2) tracking what the robot has yet to learn; (3) recalling personal habits and preferences that might otherwise be omitted; and (4) avoiding repeated explanations of information already taught across successive interactions. To decide what is worth learning, the robot should identify information relevant to subsequent household assistance and check it against existing knowledge, prioritizing useful gaps and avoiding questions about facts already taught. Without visibility into what the robot already knows, however, users may struggle to understand its learning state and decide what remains to be taught~\cite{huang2020nonverbal}. Proactive questions can make clear to users what the robot remains uncertain about~\cite{chao2010transparent} and have also been used to acquire object ownership knowledge~\cite{hashimoto2026actowl}. For example, a robot may already know who owns a cup but not where it is usually kept. It should then ask about the cup's usual storage location. Such a question prompts the user to recall a routine they might otherwise omit. The system should therefore guide teaching through questions about useful gaps in personalized knowledge, with the aim of reducing users' effort in deciding what to teach, recalling relevant information, and tracking what has already been taught.

\subsection*{D3: Make question content relevant to the user's activity context.}
For teaching during everyday activities, we identify three needs: (1) selecting among an object's knowledge gaps according to the current activity~\cite{li2025satori}; (2) making the connection between the question and the ongoing object interaction clear; and (3) enabling users to answer within the ongoing activity using immediate contextual cues, reducing the need to reconstruct the activity context afterward~\cite{beyer1998contextual,consolvo2007insitu}. Question content should therefore relate to how the user is currently using or handling an object~\cite{cai2025aiget}. Egocentric activity understanding can draw on cues such as the objects being manipulated, the actions performed, and changes in object state~\cite{grauman2022ego4d}. Interpreting the user's actions and task goals through these cues can inform the selection of proactive assistance suited to the current context~\cite{li2025satori}. We distinguish two strategies for connecting questions to activity context: activity relevance guides which missing information to ask about, and contextualized questioning guides how to phrase the question using observed object and activity cues. For example, when a user drinks water from a cup whose usual use is unknown, activity relevance favors asking about that use. Contextualized questioning makes the observed action explicit: ``You're drinking water from this cup. What do you usually drink from it?'' The question uses the current action to elicit knowledge about the user's usual use of that particular cup. Providing such context can help users understand and answer the robot's questions~\cite{rosenthal2009questions}. The system should therefore ask while the relevant object and activity cues remain available, so users can refer to the current situation rather than reconstructing it afterward.

\begin{table*}[t]
    \centering
    \caption{Design considerations for EgoAsk, summarizing the goals, descriptions, and key strategies for supporting the teaching of personalized object knowledge during everyday activities.}
    \label{tab:design-considerations}
    \small
    \renewcommand{\arraystretch}{1.2}
    \begin{tabularx}{\textwidth}{
        @{}
        >{\raggedright\arraybackslash}p{0.21\textwidth}
        >{\raggedright\arraybackslash}p{0.27\textwidth}
        >{\raggedright\arraybackslash}X
        @{}
    }
        \toprule
        \textbf{Design Consideration}
        & \textbf{Description}
        & \textbf{Key Features and Strategies} \\
        \midrule
        \textbf{D1. Enable the robot to share the user's first-person view.}
        &
        Give the robot visual access to the objects the user interacts with and their surroundings, reducing the need for users to accommodate the robot's location and field of view.
        &
        \textbullet\ Share first-person video to provide contextual cues about objects and related actions during the exchange.
        \newline
        \textbullet\ Support teaching during everyday object handling, reducing the need for additional movement or object presentation.
        \\
        \addlinespace
        \textbf{D2. Proactively ask about gaps in personalized knowledge.}
        &
        User-initiated teaching requires users to decide what is worth teaching, track knowledge gaps, and avoid omissions and repeated teaching.
        &
        \textbullet\ Identify information useful for subsequent household assistance and check existing knowledge to avoid asking about facts already taught.
        \newline
        \textbullet\ Ask specific questions that make knowledge gaps clear and prompt users to recall relevant habits and preferences.
        \\
        \addlinespace
        \textbf{D3. Make question content relevant to the user's activity context.}
        &
        Users need to understand how questions relate to the current activity and answer using immediate contextual cues, with less need to reconstruct the activity context afterward.
        &
        \textbullet\ Use activity relevance to select which knowledge gap to ask about based on the user's current object interaction.
        \newline
        \textbullet\ Use contextualized questioning to make the connection to the observed object and activity explicit in the wording.
        \newline
        \textbullet\ Ask while the relevant object and activity cues remain available for users to draw on when answering.
        \\
        \bottomrule
    \end{tabularx}
\end{table*}

While D1 gives the robot visual access to the user's object interactions and D2 guides it to ask about gaps in personalized knowledge, D3 connects these questions to the user's activity context. Table~\ref{tab:design-considerations} summarizes these design considerations and the key features and strategies associated with each. We used these considerations to design and implement EgoAsk, a system that enables robots to learn personalized object knowledge through smart glasses during everyday activities. Section~\ref{sec:system} describes how shared first-person perception, personalized knowledge discovery, and activity-relevant questioning implement these considerations.

\section{EgoAsk}
\label{sec:system}
\begingroup
\setlength{\emergencystretch}{3em}
\tolerance=2000
\renewcommand{\dbltopfraction}{0.85}
\renewcommand{\dblfloatpagefraction}{0.65}
\renewcommand{\textfraction}{0.1}
\allowdisplaybreaks[1]

We introduce EgoAsk, a prototype system built around the design considerations in Section~\ref{sec:design-considerations} to acquire personalized object knowledge through interactions embedded in everyday activities. This section presents the system overview and its key features; implementation details and agent prompts are provided in Appendices~\ref{app:system_details} and~\ref{app:agent_prompts}.

\subsection{System Overview}
\label{sec:system:overview}

EgoAsk operates through smart glasses and a robot that shares the user's first-person view. The glasses provide video, a near-eye display, and a voice interface, allowing the robot to observe objects and related activities while the user goes about everyday tasks. The system receives user and household information, including personal relationships, together with descriptions of the robot's supported capabilities. It also retrieves object knowledge and interaction history accumulated during previous exchanges. The system comprises three components for object grounding and personalized knowledge discovery; context analysis and situated question generation; and user input interpretation, knowledge update, and retrieval (Figure~\ref{fig:framework}). The third component uses a personalized knowledge graph to store and retrieve learned knowledge. Together, these components connect shared visual access (D1), proactive knowledge acquisition (D2), and activity-relevant questioning (D3).

The system supports two interaction paths. For robot-initiated questioning, it identifies an object of interest, discovers candidate knowledge from potential assistance scenarios, filters known or deferred needs, and selects a valuable gap using the current activity before generating a question. For user-initiated teaching, the user's utterance enters input interpretation and knowledge update directly, using the visible objects and dialogue context to resolve its referent. Both paths store user-provided knowledge for subsequent assistance and feed it back into memory filtering to reduce repeated questions.

\begin{figure*}[t]
  \centering
  \includegraphics[width=0.985\linewidth]{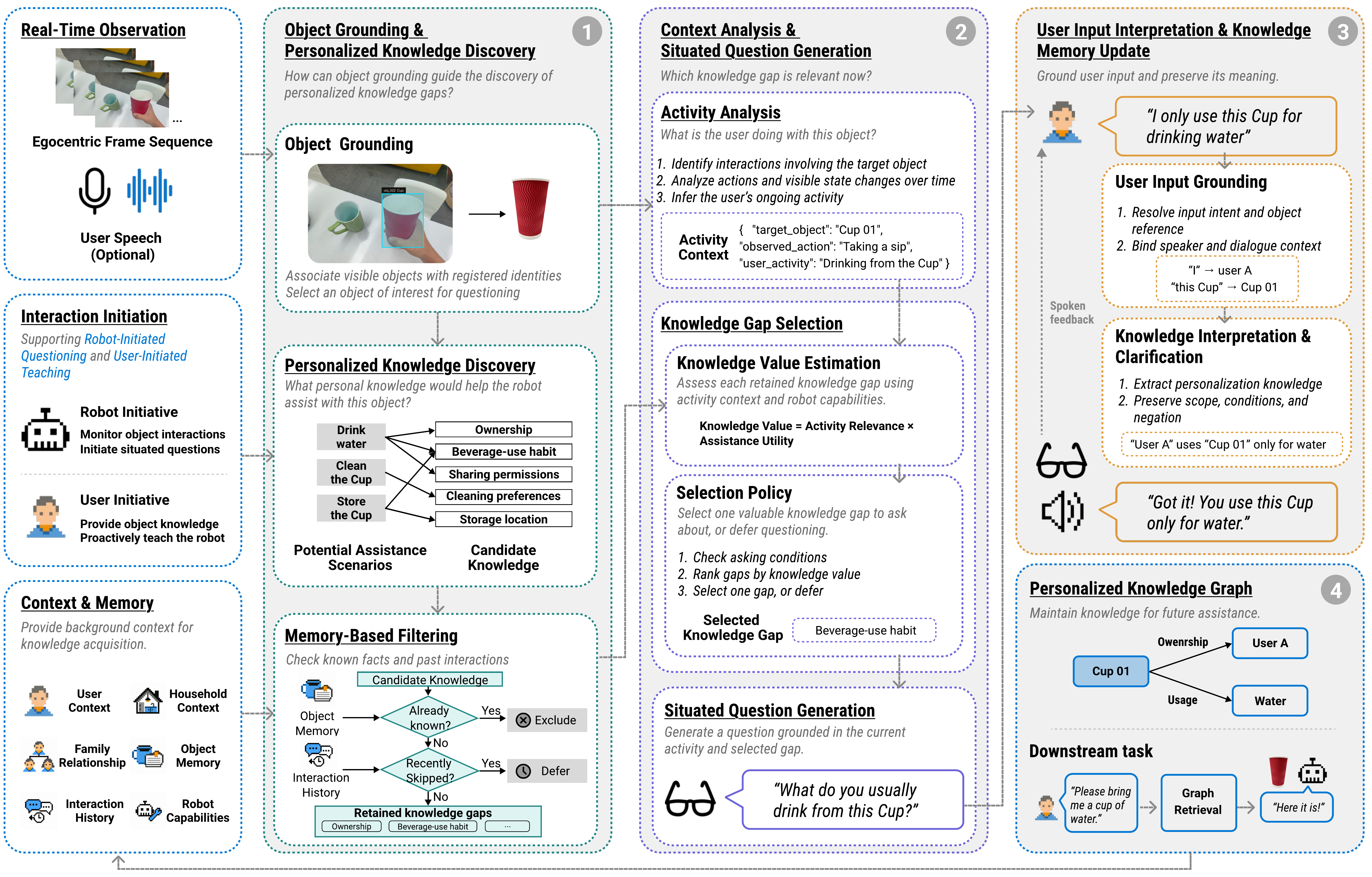}
  \caption{EgoAsk's framework for acquiring and reusing personalized object knowledge. Shared first-person observations, optional user speech, and context and memory support both robot-initiated questioning and user-initiated teaching. \textbf{(1) Object grounding and personalized knowledge discovery} associates the object of interest with a persistent identity and derives candidate knowledge needs from potential assistance scenarios. One scenario can motivate multiple needs, such as ownership, beverage-use habits, and sharing permissions. Memory-based filtering removes answered needs and defers recently skipped ones. \textbf{(2) Context analysis and situated question generation} interprets the user's activity involving the target object, ranks retained gaps by the product of Activity Relevance and Assistance Utility, and selects one gap or defers questioning. In the illustrated example, drinking from Cup~01 motivates the question, ``What do you usually drink from this cup?'' \textbf{(3) User input interpretation, knowledge update, and retrieval} resolves references in the response, preserves its user scope and qualifiers, and updates a personalized knowledge graph. ``I only use this cup for drinking water'' becomes a usage statement about User~A and Cup~01, with spoken feedback communicating the interpretation. The graph supports subsequent acquisition and task-related retrieval within this component, illustrated by selecting a cup for a later water request. User-initiated teaching enters input interpretation directly, without requiring a robot-generated question.}
  \Description{The framework shows object grounding and personalized knowledge discovery, activity analysis and situated question generation, and user input interpretation and knowledge update, alongside the personalized knowledge graph used for storage and retrieval. Observation, interaction initiation, and context appear on the left. The example follows a cup-related question and user answer through knowledge acquisition to later retrieval for a water request.}
  \label{fig:framework}
\end{figure*}

\subsection{Key Features of the EgoAsk System}
\label{sec:system:key_features}

Table~\ref{tab:prompt_design} summarizes the prompt and context design strategies used to discover, prioritize, and ask about personalized object knowledge.
\begin{table*}[t]
  \caption{Prompt and context design strategies for personalized knowledge acquisition in EgoAsk. Examples illustrate the intended behavior; agent prompts and output requirements appear in Appendix~\ref{app:agent_prompts}.}
  \label{tab:prompt_design}
  \small
  \setlength{\tabcolsep}{5pt}
  \renewcommand{\arraystretch}{1.15}
  \begin{tabularx}{\textwidth}{@{}>{\raggedright\arraybackslash}p{0.18\textwidth}>{\raggedright\arraybackslash}p{0.25\textwidth}>{\raggedright\arraybackslash}X@{}}
    \toprule
    \textbf{Module} & \textbf{Challenge} & \textbf{Strategies and examples} \\
    \midrule
    Personalized Knowledge Discovery
    & \textbf{Identify personal knowledge needs.} Determine what to learn beyond generic object properties and predefined attributes.
    & \textbf{Derive needs from assistance scenarios.} Supply relevant BEHAVIOR-derived tasks within the robot's capabilities. Ask which personal knowledge could change an assistance decision, and derive descriptive labels. Selecting a cup for a drink can yield beverage-use habits and sharing permissions. \\
    \addlinespace[5pt]
    Knowledge Gap Selection
    & \textbf{Prioritize useful unknowns.} An unanswered need may be unrelated to the current activity or have little assistance value.
    & \textbf{Score relevance and utility.} Provide activity context, retained gaps, and robot capabilities. Ask the LLM to score Activity Relevance and Assistance Utility using explicit criteria; rank their product and allow deferral. Drinking from a cup makes beverage-use habits relevant to selecting a cup for a later drink request. \\
    \addlinespace[5pt]
    Situated Question Generation
    & \textbf{Ask a grounded, answerable question.} Refer clearly to the object and selected need without presupposing a personal fact.
    & \textbf{Constrain wording with context and examples.} Supply the target, activity, selected gap, and relevant memory, with few-shot demonstrations. Ask about one need at a time: ``What do you usually drink from this cup?'' rather than ``Why do you only use it for water?'' \\
    \bottomrule
  \end{tabularx}
  \Description{3 rows summarize EgoAsk's design strategies: assistance scenarios guide personalized knowledge discovery, activity relevance and assistance utility guide gap selection, and contextual inputs with few-shot examples guide question wording.}
\end{table*}

\subsubsection{Object Grounding and Personalized Knowledge Discovery (D1, D2)}
\label{sec:system:knowledge_discovery}

\paragraph{Object Grounding.}
EgoAsk identifies an object of interest in the shared first-person view and associates it with a persistent instance identity as the target of proactive questioning (D1). It uses YOLOE to detect objects~\cite{wang2025yoloe}, ByteTrack to associate detections across frames~\cite{zhang2022bytetrack}, and DINOv2 to extract instance features~\cite{oquab2023dinov2}. Feature matching connects repeated observations to online instance records, with new records created for persistently observed new objects. Among objects with sustained visibility and stable identity associations, the system selects the largest bounding box as the object of interest. The robot then discovers personalized knowledge about this object and analyzes the user's activity involving it. Detection, online identity maintenance, and selection rules are detailed in Appendix~\ref{app:grounding}.

\paragraph{Personalized Knowledge Discovery.}
To identify personal knowledge that could help the robot assist with the target object (D2), EgoAsk first retrieves household tasks in which the robot may later use or handle it. The system uses the object category to retrieve tasks from a catalog derived from BEHAVIOR activity definitions~\cite{srivastava2022behavior,li2023behavior1k}, then filters them by the robot's supported capabilities. This produces potential assistance scenarios, such as selecting a cup for water, cleaning it, or putting it away; catalog construction and retrieval are described in Appendix~\ref{app:assistance_catalog}.

Given these scenarios, an LLM-based knowledge discovery agent identifies personalized knowledge that the robot needs to learn from the user. For each scenario, the agent identifies personalized information that could affect an assistance decision. For each information need, the prompt asks the agent to propose 2 hypothetical answers and explain how each would affect the robot's choice of object or operation. It then derives descriptive labels and returns the information needed, associated scenarios, and affected decisions (Appendix~\ref{app:discovery_candidates}). For example, selecting a cup to fill with water may require knowing who owns it, which beverages it is usually used for, and whether others may use it, yielding ownership, beverage-use habit, and sharing-permission candidates. One scenario can motivate several needs, and the same need can support several scenarios. These labels describe what to learn; they do not supply presumed personal answers.

\paragraph{Memory-Based Filtering.}
To reduce repeated questioning (D2), EgoAsk compares candidates with existing knowledge and interaction history. The system retrieves memory by user and object identity, and an LLM checks whether the retrieved knowledge answers each candidate need. If a need is only partially answered, the system retains the unanswered part. Known needs are excluded, while needs recently skipped or declined are deferred. Retained knowledge gaps proceed to context analysis and question generation; the prompt and filtering rules appear in Appendix~\ref{app:discovery_memory_filtering}.

\subsubsection{Context Analysis and Situated Question Generation (D2, D3)}
\label{sec:system:situated_questioning}

\paragraph{Activity Analysis.}
To connect question content to the user's current object-related activity (D3), EgoAsk analyzes how the user is using or handling the target. The system marks the target object in recent first-person frames and provides the frames to a vision-language model (VLM) in temporal order. The prompt directs attention to actions involving the object~\cite{damen2022rescaling}, visible state changes during interaction~\cite{grauman2022ego4d}, and other participating objects to infer the ongoing activity; input settings and prompts appear in Appendix~\ref{app:activity_prompt}. It separates observed actions from activity interpretations: for example, lifting a cup to the mouth, taking a sip, and lowering it support an interpretation of drinking from the cup. The output describes the user's observed actions involving the target object and the inferred activity, leaving unsupported interpretations unspecified.

\paragraph{Knowledge Gap Selection.}
Knowledge discovery identifies potentially useful information needs for future assistance; knowledge gap selection determines which retained gap is worth asking about in the current activity (D2, D3). EgoAsk makes this choice by ranking the retained gaps by knowledge value. The LLM receives the activity context, candidate gaps and their associated assistance scenarios, and supported robot capabilities. Using explicit scoring criteria, as in rubric-based LLM evaluation~\cite{liu2023geval}, it assigns each gap an Activity Relevance score and an Assistance Utility score, with a brief basis for each. Both use a 0--2 scale: Activity Relevance denotes no clear, indirect, or direct connection to the activity; Assistance Utility denotes no clear use, supplementary information, or a direct contribution to object selection or an operation. The application computes the heuristic knowledge value as
\begin{equation}
  V(g_i) = R(g_i) \times U(g_i),
  \label{eq:knowledge_value}
\end{equation}
where $g_i$ denotes the $i$-th retained knowledge gap, $R(g_i)$ is its Activity Relevance score, $U(g_i)$ is its Assistance Utility score, and $V(g_i)$ is the resulting Knowledge Value. The scoring prompt and selection rules appear in Appendix~\ref{app:gap_selection}. For example, when the user is drinking from a cup, beverage-use habits are directly relevant and can inform whether the robot selects that cup for a requested drink. When dialogue conditions permit, the system selects the highest-scoring candidate with a positive value; otherwise, it defers questioning.

\paragraph{Situated Question Generation and Delivery.}
To help users understand and answer questions within their current activity (D3), EgoAsk turns the selected knowledge gap into a question about the target object. Related embodied-agent and vision-language research uses questions to obtain task-relevant information or clarify ambiguity~\cite{shen2025elba,jian2025teaching}. EgoAsk supplies the LLM with the object, activity, selected need, relevant memory, and recent dialogue, together with few-shot examples of situated question wording (Appendix~\ref{app:question_generation}). The prompt asks for one knowledge need at a time, a clear object reference, and no presupposition of unconfirmed facts. For example, when the user is drinking from a cup and the selected need concerns beverage-use habits, the question is ``What do you usually drink from this cup?'' The glasses present the question through speech and visually indicate the target object on the near-eye display. Users can answer or skip; the system retains the question's object and candidate associations for input interpretation.

\subsubsection{User Input Interpretation, Knowledge Update, and Retrieval}
\label{sec:system:input_interpretation}

EgoAsk uses an LLM to transform question--answer exchanges and unsolicited teaching into personalized knowledge. The model resolves references using the current question, target object, and dialogue context, then extracts a structured statement identifying whom the knowledge concerns and the conditions and restrictions under which it applies. The system uses the Neo4j graph database\footnote{\url{https://neo4j.com/}} to store relationships among users, objects, and personal knowledge for updates and later retrieval; prior personalized kitchen assistance has also used this database to organize knowledge~\cite{guan2023robot}. For example, ``I only use this cup for water'' becomes a usage statement associated with User~A, Cup~01, and Water, retaining the exclusivity qualifier. The application validates the record, stores its source and acquisition time, merges duplicate knowledge records, and updates existing statements when the user explicitly corrects them. Unclear input prompts clarification.

Learned knowledge supports object retrieval and selection for subsequent tasks. Given a request, the LLM extracts the required object category, intended use, and user constraints. The system retrieves relevant knowledge records that have not been superseded by user corrections and returns candidate object instances. For example, when User~A asks, ``Please bring me a cup of water,'' their recorded beverage-use habit can identify Cup~01 as a candidate; current perception then establishes its location and availability. This allows knowledge acquired during teaching to inform later assistance without requiring the user to repeat the same preference.

\endgroup

\section{User Study}
\label{sec:study}

EgoAsk combines robot-initiated questioning with interaction grounded in the user's current activity, enabling the robot to ask about missing personalized knowledge while the user handles objects. Robot-initiated questions can make the robot's information needs visible, helping users determine what remains to be taught~\cite{cakmak2010designing}. At the same time, asking during object interactions allows users to draw on the current context when answering, potentially reducing the need to recall and reconstruct that context afterward~\cite{smith2001environmental}. However, while this interaction may make teaching easier, it may also interrupt the user's ongoing activity~\cite{adamczyk2004not}. How proactive, situated questioning changes the burden of teaching therefore remains an empirical question.

To examine this question, we conducted a mixed-methods within-subjects study comparing User-Led Teaching, Post-Task Questioning, and In-Task Questioning (EgoAsk). We investigated how teaching initiative and question timing redistribute users' subjective teaching burden through the following research questions:

\begin{itemize}
  \item \textbf{RQ1:} How does robot-initiated questioning affect users' gap-monitoring burden compared with User-Led Teaching?
  \item \textbf{RQ2a:} How does asking questions during the task affect users' context-reconstruction burden compared with asking after the task?
  \item \textbf{RQ2b:} How does asking questions during the task affect users' interruption burden compared with asking after the task?
\end{itemize}

We test three directional hypotheses. Robot-initiated questioning reduces gap-monitoring burden compared with User-Led Teaching (H1). EgoAsk's in-task questioning reduces context-reconstruction burden compared with Post-Task Questioning (H2a). EgoAsk's in-task questioning increases interruption burden compared with Post-Task Questioning (H2b).

\subsection{Task and Study Design}
\label{sec:study:design}
\begin{figure*}[t]
  \centering
  \includegraphics[width=\linewidth]{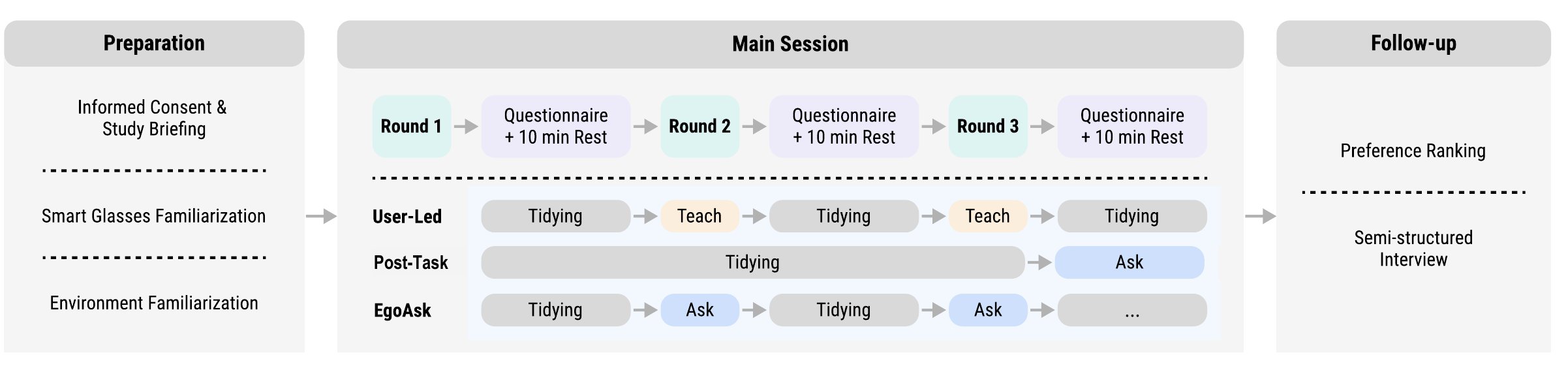}
  \caption{User study procedure: preparation, three teaching rounds, and preference ranking followed by an interview. Each round was followed by a questionnaire and a ten-minute rest. The condition timelines illustrate User-Led teaching and EgoAsk questioning during tidying, and Post-Task questioning after tidying. Condition order was counterbalanced across participants.}
  \Description{The diagram shows preparation, a main session with three rounds, and a follow-up with preference ranking and a semi-structured interview. Preparation includes informed consent, study briefing, smart glasses familiarization, and environment familiarization. Each round uses one teaching condition and is followed by a questionnaire and a ten-minute rest. The three condition timelines show user-initiated teaching interleaved with tidying, robot questioning after tidying, and EgoAsk questioning interleaved with tidying. The displayed condition rows do not indicate the order of rounds.}
  \label{fig:study-design}
\end{figure*}

We used a within-subjects design comparing three teaching paradigms: User-Led Teaching, Post-Task Questioning, and In-Task Questioning (EgoAsk), as summarized in Table~\ref{tab:conditions}. For compact reporting in tables and figures, we use the labels User-Led, Post-Task, and EgoAsk. All conditions used the same smart glasses and EgoAsk backend. The experimental manipulation focused on who initiated the exchange and when the user answered.

The study took place in a laboratory arranged as a simulated home environment, with areas corresponding to a desk, a coffee table, a bar counter, and a cabinet. Participants completed three household tidying rounds using three sets of five everyday personal objects. At the start of each round, the experimenter provided a box containing one object set; the full object lists appear in Appendix~\ref{app:objects}. Participants wore the smart glasses, took objects out of the box, and placed them wherever they thought appropriate.

To support comparability across the three teaching paradigms, we used a common household tidying task to reduce variation in activity demands and task structure, focusing the comparison on subjective burdens associated with teaching initiative and question timing. Handling, moving, and putting away objects naturally brought them into the first-person view and provided object, action, and location cues for situated questioning. Tidying directly involves personalized object knowledge, such as who owns something, how it is used, and where it belongs, all of which are relevant to household robot assistance~\cite{wu2023tidybot,kapelyukh2022myhouse,hashimoto2026actowl,ikeuchi2024semantic}. These facts cannot be reliably inferred from appearance, which makes them appropriate teaching content for this study.

\begin{table}[h]
\caption{The three teaching paradigms compared in the study.}
\label{tab:conditions}
\centering
\begin{tabularx}{\linewidth}{@{}>{\raggedright\arraybackslash}p{0.29\linewidth}>{\raggedright\arraybackslash}p{0.19\linewidth}X@{}}
\toprule
Condition & Initiator & Timing \\
\midrule
User-Led          & User  & During activity \\
Post-Task         & Robot & After activity  \\
EgoAsk            & Robot & During activity \\
\bottomrule
\end{tabularx}
\end{table}

In the User-Led Teaching condition, the robot did not ask questions, and participants decided what information to provide, when to provide it, and whether to add further information. In the Post-Task Questioning condition, participants first completed the tidying task, then moved away from the table to answer a batch of robot questions. The objects remained where participants had placed them. In the In-Task Questioning condition, EgoAsk used the current object, activity context, and retained knowledge gaps to select and generate situated questions during tidying. Within each condition, participants could tidy and interact with the smart glasses at their own pace.

Both robot-initiated conditions used EgoAsk's Personalized Knowledge Discovery module and memory-based filtering to determine which knowledge gaps to ask about (Section~\ref{sec:system:knowledge_discovery}). We matched the information sought between these conditions and applied the same limit of at most three questions per object. The wording of individual questions was generated in real time from the ongoing dialogue and could therefore differ between conditions. Appendix~\ref{app:study_knowledge_targets} provides illustrative knowledge candidates for study objects.

We counterbalanced both condition order and object-set assignment across participants using Latin-square rotations. The 18 participants were distributed across three balanced sequences, with six participants per sequence. Thus, each teaching paradigm appeared equally often in each round position, and each object set appeared under different paradigms across participants.

\subsection{Procedure}
\label{sec:study:procedure}

Each session lasted approximately 60 minutes and followed three phases (Figure~\ref{fig:study-design}). In Phase 1 (Preparation), participants completed informed consent and received a briefing on the household tidying task and the goal of teaching the robot personalized object knowledge to support future assistance. The experimenter then fitted the glasses and led a brief familiarization covering the HUD display, the tap-to-record interaction, and the simulated home layout. During familiarization, we used narrative framing to invite participants to treat the simulated environment as their home and the study objects as household belongings~\cite{syrdal2015integrating}. Participants were asked to arrange the objects and provide information to the robot according to their own preferences.

In Phase 2 (Teaching Rounds), participants completed three rounds in succession. Before each round, participants inspected the five objects in the box for that round. Participants then wore the glasses and tidied the objects in the simulated home. The tidying phase ended when all five objects had been placed. In Post-Task Questioning, the subsequent question-and-answer phase ended once the robot had asked all questions for that round. After each round, participants independently completed the post-round questionnaire and rested for ten minutes before proceeding to the next round or, after the final round, the interview phase.

In Phase 3 (Interview), participants ranked the three teaching paradigms from most to least preferred and took part in a semi-structured interview.

\subsection{Instruments}
\label{sec:study:instruments}

\textbf{Questionnaire.} After each round, participants completed the same questionnaire across all three teaching conditions to assess overall workload and teaching-specific burdens. The questionnaire included the six NASA-TLX dimensions on seven-point scales~\cite{hart1988tlx}. To distinguish sources of teaching burden beyond overall workload, we supplemented these with nine study-specific seven-point Likert items, organized into three groups of three. The first group assessed the effort and concern involved in judging what information the robot still lacked and supplying additional information~\cite{cakmak2010designing}. The second assessed difficulties in recalling the circumstances of handling an object, ensuring that the information provided referred to the correct object, and providing accurate information~\cite{smith2001environmental,clark1986referring}. The third assessed whether interactions disrupted tidying, occurred at abrupt moments, or required effort to resume the task~\cite{adamczyk2004not,bailey2008workload}. The questionnaire ended with an optional open-ended question inviting participants to describe anything particularly memorable, confusing, uncomfortable, or smooth in that round. All questionnaires were completed independently and anonymized, and the full item list appears in Appendix~\ref{app:questionnaire}. Internal consistency was assessed before computing scale scores, and the full statistical procedure is reported with the quantitative results in Section~\ref{sec:results}.

\textbf{Interview.} After all three rounds, participants ranked the three paradigms by overall preference and completed a semi-structured interview. Participants explained their preferences and reflected on overall workload, gap-monitoring burden, context-reconstruction burden, and interruption burden. We also asked how they experienced robot-initiated questions during versus after tidying, whether the robot's feedback adequately confirmed what they had taught, and what concerns they had about using the system at home. The full interview guide appears in Appendix~\ref{app:interview}. We analyzed interview transcripts and open-ended questionnaire comments using codebook thematic analysis~\cite{braun2006thematic}, with the coding approach described in Section~\ref{sec:results:qualitative}.

\subsection{Materials and Tools}
\label{sec:study:materials}

Participants wore RayNeo X3 Pro smart glasses throughout the study. The glasses integrate a first-person camera, a near-eye display, a microphone, a speaker, and temple-tap input. The HUD displayed bounding boxes and class labels for detected objects, using color to indicate whether an object was currently available for teaching. A status bar displayed the robot's current question and whether audio was being recorded. Participants heard the robot's questions and confirmations through the glasses' speaker. Participants tapped the right temple once to start recording, provided teaching information or answered a robot question, and tapped again to stop. A language model converted the transcribed speech into structured knowledge statements, retaining whom the information concerned and when it applied. The system used these statements to update the corresponding object's knowledge in memory; the record checks and update procedure are described in Section~\ref{sec:system:input_interpretation}.

A wheeled dual-arm robot was present in the study environment, equipped with a touchscreen and an Intel RealSense D435 camera. The robot was presented as the household agent that learns personalized object knowledge. Its touchscreen provided a natural-language retrieval interface that displayed the object identified in response to a user's request. This retrieval functionality was not included in the user study.

\begin{figure*}[t]
  \centering
  \includegraphics[width=\linewidth]{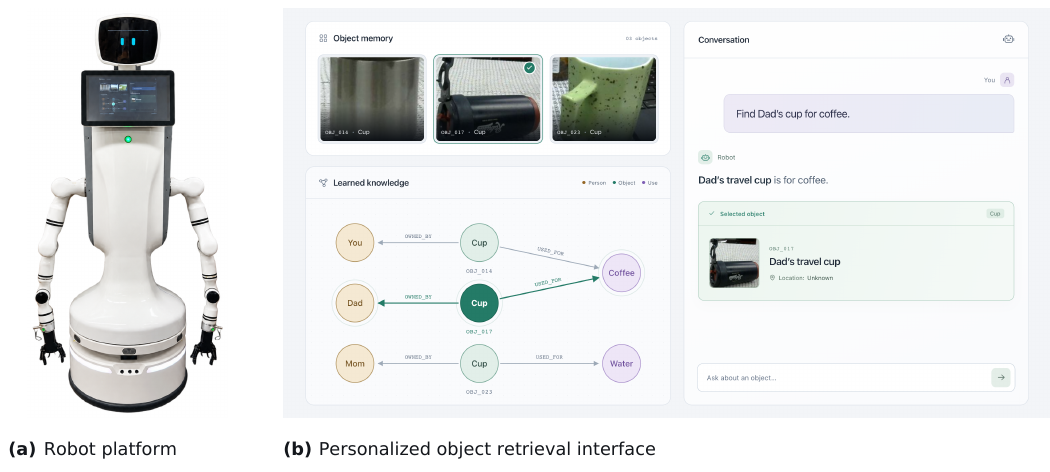}
  \caption{Robot platform and illustrative personalized object retrieval interface. (a) The wheeled dual-arm robot with a head display and a torso-mounted touchscreen. (b) An example interface showing identified cup instances, their ownership and usage relations, and the selection of Dad's cup in response to ``Find Dad's cup for coffee.'' The interaction was constructed for demonstration purposes. The selected object's current location is shown as unknown.}
  \Description{Two panels appear side by side. The left panel shows the full robot, including its head display, touchscreen, two arms, grippers, and mobile base. The right panel shows three identified cup instances and a knowledge graph linking them to You, Dad, and Mom and to coffee or water. Dad's cup and its ownership and coffee-use relations are highlighted. A dialogue asks for Dad's cup for coffee, and a result card identifies object OBJ 017 with its location marked Unknown.}
  \label{fig:robot-retrieval}
\end{figure*}

Figure~\ref{fig:robot-retrieval} shows the robot platform and an illustrative retrieval interface demonstrating how the robot uses learned ownership and usage knowledge to identify the specific object referred to in a user's request.

\begin{figure}
  \centering
  \includegraphics[width=0.8\linewidth]{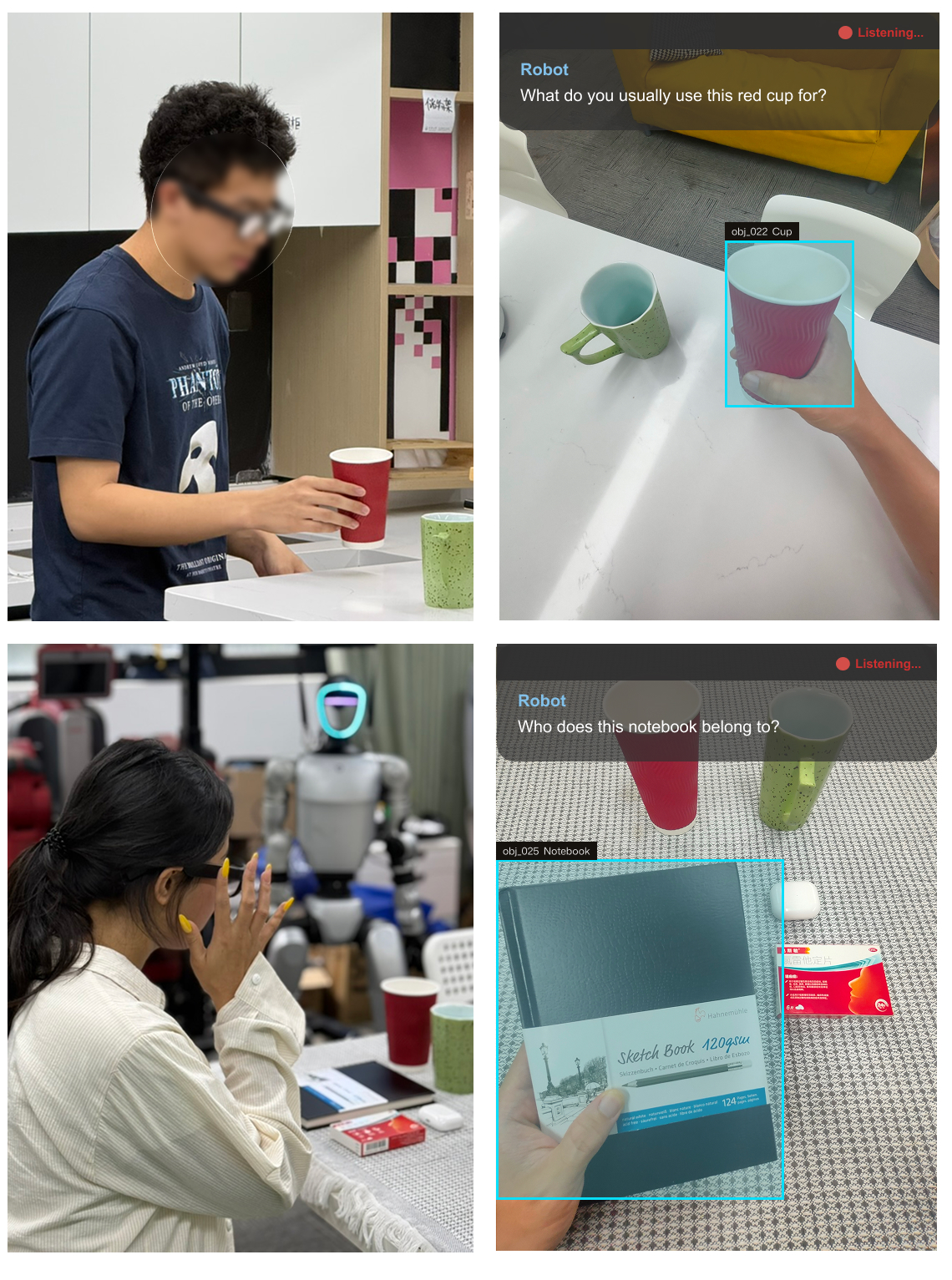}
  \caption{Study setting and illustrative egocentric teaching views. Left: participants wearing smart glasses while organizing everyday objects during the household tidying task. Right: corresponding first-person views with illustrative overlays showing the target object's bounding box, persistent ID and category, the robot's question, and the listening state. The examples illustrate how the robot identifies a target object with missing personalized knowledge and initiates a question about it, asking about the red cup's use (top) and the notebook's ownership (bottom).}
  \Description{Four panels in two rows. The left panels show participants wearing smart glasses, with their faces blurred. The upper-right panel shows a held red cup labeled obj\_022 Cup and the question, What do you usually use this red cup for? The lower-right panel shows a held notebook labeled obj\_025 Notebook and the question, Who does this notebook belong to? Both first-person views include a cyan target bounding box and a red Listening indicator. The interface overlays are illustrative.}
  \label{fig:live}
\end{figure}

The glasses streamed egocentric video to the study server, which handled perception, dialogue coordination, knowledge storage, and speech processing. EgoAsk's language-model components used GPT-5 (\texttt{gpt-5-2025-08-07}) for activity analysis, knowledge discovery and filtering, question selection and generation, input interpretation, and subsequent object retrieval, as described in Section~\ref{sec:system}.

\subsection{Participants and Ethics}
\label{sec:study:participants}

We recruited 18 participants through laboratory convenience sampling for a study on teaching a household robot about personal objects with smart glasses. Participants (ten male, eight female) were aged 19--33 years ($M = 23.2$, $SD = 3.2$). Educational backgrounds ranged across undergraduate ($n = 14$), master's ($n = 2$), and doctoral ($n = 2$) programs. All participants reported normal hearing. Of the participants, 17 reported normal or corrected-to-normal vision; the remaining participant had mild myopia that did not interfere with viewing information on the HUD. All wore the glasses comfortably and reported no history of AR or VR motion sickness. Most had prior experience with smart devices, robotic systems, or head-mounted displays.

The study was reviewed and approved by the relevant institutional ethics review committee prior to data collection. All participants provided informed consent, participated voluntarily, and could withdraw from the study at any time. Participant data were anonymized for analysis and reporting. First-person video and audio were recorded for research analysis. Participants received \$10 compensation upon completion.

\section{Results}
\label{sec:results}

This section reports the analysis procedure and results for the post-round questionnaire, the preference ranking, and the semi-structured interview. The questionnaire results test the 3 research questions, the preference ranking supplements overall acceptability, and the thematic analysis explains the subjective mechanisms behind the questionnaire patterns. We report questionnaire and preference results first (Section~\ref{sec:results:quantitative}), followed by the thematic analysis (Section~\ref{sec:results:qualitative}).

\subsection{Questionnaire and Preference Results}
\label{sec:results:quantitative}

We compared questionnaire ratings across the 3 teaching paradigms using Friedman tests~\cite{friedman1937ranks} with paired data from 18 participants. The questionnaire used 7-point ratings. Exploratory pairwise comparisons used two-sided Wilcoxon signed-rank tests~\cite{wilcoxon1945individual}, with $p$-values and effect sizes $r$ reported.

Overall workload was measured by averaging the 6 NASA-TLX dimensions~\cite{hart1988tlx}, after reverse-scoring the Performance item (Q4) as $8-Q4$ so that higher scores indicate greater workload. Gap-monitoring burden and interruption burden were each calculated as the mean of their 3 corresponding items. We assessed internal consistency using Cronbach's $\alpha$ on responses pooled across conditions. As shown in Table~\ref{tab:burden}, $\alpha$ was .81 for NASA-TLX, .84 for gap-monitoring burden, and .77 for interruption burden.

The study-specific item set intended to assess context-reconstruction burden (Q10--Q12) showed low pooled internal consistency ($\alpha=.28$; mean inter-item $r=.10$). We therefore analyzed the 3 items separately without computing a composite score. Q10 measures the reported need to reconstruct the handling context, Q11 measures object-confirmation effort, and Q12 measures difficulty providing accurate object information. Complete item-level results appear in Appendix~\ref{app:questionnaire}.

\begin{figure}[t]
  \centering
  \includegraphics[width=\linewidth]{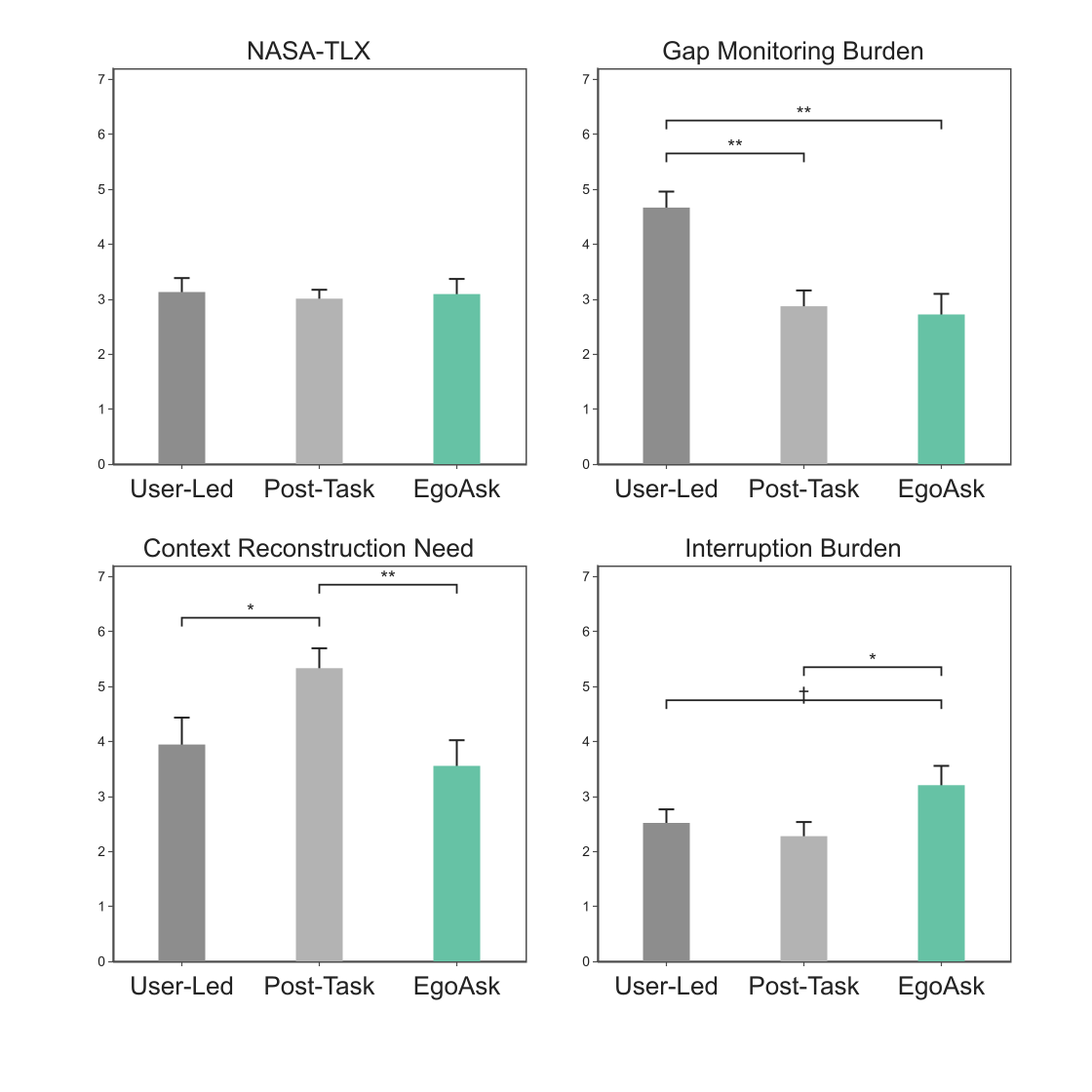}
  \caption{Mean ratings by condition across 4 questionnaire measures, including NASA-TLX. Context reconstruction is represented by the single item Q10, not a Q10--Q12 composite. Error bars show $\pm 1$ SEM. Brackets indicate exploratory pairwise Wilcoxon comparisons: * $p<.05$, ** $p<.01$, $\dagger$ $p<.10$. The interruption omnibus test was not significant ($p=.096$).}
  \label{fig:workload}
\end{figure}

\textbf{Overall burden pattern.} Overall workload did not differ significantly across conditions (NASA-TLX; Friedman $\chi^2(2)=1.21$, $p=.546$), with means of 3.13, 3.01, and 3.09 for User-Led Teaching, Post-Task Questioning, and EgoAsk, respectively (Table~\ref{tab:burden}). The teaching-specific measures showed different descriptive patterns. Gap-monitoring burden was highest in User-Led Teaching, context-reconstruction need (Q10) was highest in Post-Task Questioning, and interruption burden was numerically highest in EgoAsk (Figure~\ref{fig:workload}).

\begin{table*}[t]
\caption{Descriptive statistics for the 4 questionnaire measures across conditions ($N=18$; 1--7 scale; higher scores indicate greater burden or context-reconstruction need). Bold marks the highest mean among conditions for each teaching-specific measure. Alpha estimates pool responses across conditions.}
\label{tab:burden}
\centering
\begin{tabularx}{0.8\linewidth}{@{}>{\raggedright\arraybackslash}Xrrrr@{}}
\toprule
Measure & User-Led $M$($SD$) & Post-Task $M$($SD$) & EgoAsk $M$($SD$) & $\alpha$ \\
\midrule
NASA-TLX                      & 3.13 (1.08) & 3.01 (0.68) & 3.09 (1.17) & .81 \\
Gap-monitoring burden          & \textbf{4.67 (1.23)} & 2.87 (1.22) & 2.72 (1.59) & .84 \\
Context-reconstruction need (Q10)   & 3.94 (2.07) & \textbf{5.33 (1.53)} & 3.56 (1.98) & n/a \\
Interruption burden            & 2.52 (1.04) & 2.28 (1.09) & \textbf{3.20 (1.50)} & .77 \\
\bottomrule
\end{tabularx}
\end{table*}

\textbf{RQ1: Gap-monitoring burden.} Participants reported lower gap-monitoring burden in both robot-initiated conditions than in User-Led Teaching. Scores differed across conditions (Friedman $\chi^2(2)=11.58$, $p=.003$). User-Led Teaching ($M=4.67$) was higher than Post-Task Questioning ($M=2.87$, $p=.002$, $r=.74$) and EgoAsk ($M=2.72$, $p=.007$, $r=.64$). No significant difference was detected between Post-Task Questioning and EgoAsk ($p=.648$). Individual item ratings followed the same pattern. Participants reported less need to continually assess missing information, worry about omissions, and proactively supply or confirm information in both robot-initiated conditions than in User-Led Teaching (Q7--Q9; all $p<.05$; Table~\ref{tab:context-items}). These results support H1, associating robot-initiated questioning with lower perceived monitoring burden.

\textbf{RQ2a: Context-reconstruction burden.} Participants reported a greater need to reconstruct the handling context in Post-Task Questioning on Q10. Ratings differed across conditions (Friedman $\chi^2(2) = 11.33$, $p = .003$). Post-Task Questioning ($M = 5.33$) was higher than EgoAsk ($M = 3.56$, $p = .004$, $r = .68$) and User-Led Teaching ($M = 3.94$, $p = .018$, $r = .56$). No significant difference was detected between User-Led Teaching and EgoAsk ($p = .204$).

Object-confirmation effort (Q11) had mean ratings of 3.39, 3.28, and 3.00 in User-Led Teaching, Post-Task Questioning, and EgoAsk, respectively, with no significant omnibus difference ($\chi^2(2)=1.11$, $p=.575$) or pairwise comparison. Answering difficulty (Q12) ratings clustered at the low end (78\% of responses $\leq 2$; respective means 2.39, 1.83, and 1.61). Its omnibus test was not significant ($\chi^2(2)=4.92$, $p=.085$). An exploratory comparison found higher answering-difficulty ratings in User-Led Teaching than in EgoAsk ($p=.016$, $r=.57$), but no significant difference between Post-Task Questioning and EgoAsk ($p=.271$). The Q10 result provides preliminary support for H2a, indicating a lower reported need for context reconstruction in EgoAsk than in Post-Task Questioning.

\textbf{RQ2b: Interruption burden.} EgoAsk had the highest mean interruption score ($M = 3.20$), followed by User-Led Teaching ($M = 2.52$) and Post-Task Questioning ($M = 2.28$). The omnibus test did not reach significance (Friedman $\chi^2(2) = 4.69$, $p = .096$). The exploratory comparison between EgoAsk and Post-Task Questioning reached the nominal .05 level ($p = .046$, $r = .47$), whereas the comparison with User-Led Teaching did not ($p = .087$, $r = .40$). The pairwise result provides tentative evidence in the direction of H2b, without a significant omnibus effect.

\begin{figure}
  \centering
  \includegraphics[width=\linewidth]{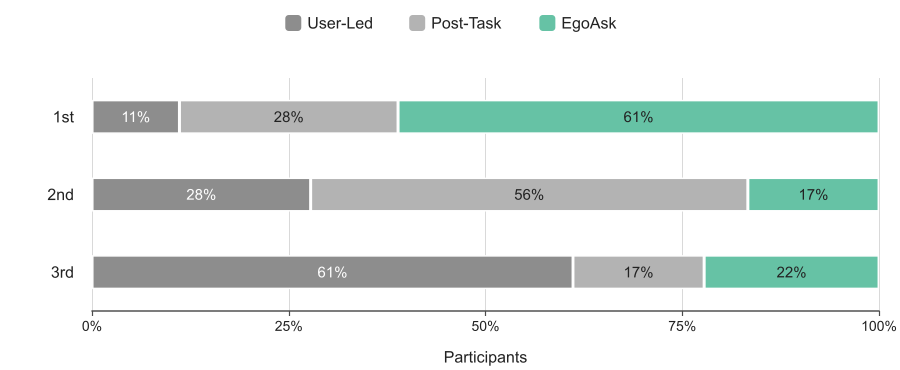}
  \caption{Distribution of overall preference rankings for the 3 teaching paradigms. Bars show the percentage of participants who placed each paradigm in each rank position.}
  \label{fig:preferences}
\end{figure}

\textbf{Overall preference.} After completing all 3 rounds, participants ranked the paradigms from most to least preferred (Figure~\ref{fig:preferences}). EgoAsk received 11 first-place rankings (61\%), followed by Post-Task Questioning with 5 (28\%) and User-Led Teaching with 2 (11\%). Post-Task Questioning was most frequently ranked second (10 participants, 56\%), while User-Led Teaching was most frequently ranked last (11 participants, 61\%).

\begin{samepage}
\subsection{Thematic Analysis}
\label{sec:results:qualitative}

To contextualize the questionnaire and preference results, we conducted codebook thematic analysis of 18 post-session semi-structured interview transcripts and the block questionnaire's open-ended responses~\cite{braun2006thematic,mcdonald2019reliability}. We developed the initial codebook from the research questions and questionnaire constructs, then refined it through pilot coding of the transcripts. Two coders independently coded all 18 interview transcripts, with Cohen's $\kappa=.897$ before disagreements were resolved through discussion~\cite{cohen1960coefficient}. The final codebook comprised four themes and eleven codes (Table~\ref{tab:themes}). Quotations were translated from Chinese with minor removal of disfluencies.

\end{samepage}

\begin{table}[t]
\caption{Themes and codes from the thematic analysis.}
\label{tab:themes}
\centering
\begin{tabularx}{\linewidth}{>{\raggedright\arraybackslash}p{\dimexpr0.48\linewidth-2\tabcolsep\relax}>{\raggedright\arraybackslash}X}
\toprule
Themes & Codes \\
\midrule
Gap-Monitoring Responsibility
  & Uncertainty about information needs \newline
    Concern about knowledge omissions \newline
    Robot-surfaced knowledge gaps \\
\addlinespace
Situational Cue Availability
  & Object cues during handling \newline
    Post-task context reconstruction \\
\addlinespace
Experiences of Question Timing
  & Conversational immediacy \newline
    Task-flow interruption \newline
    Post-task answering comfort \\
\addlinespace
Control, Correction, and Adoption Conditions
  & Confirmation and correction needs \newline
    Selective or deferrable questioning \newline
    Trustworthiness and deployment concerns \\
\bottomrule
\end{tabularx}
\end{table}

\textbf{RQ1: Gap-monitoring responsibility.} \emph{Uncertainty about information needs.} Participants described difficulty deciding what information the robot needed in User-Led Teaching. P04 did not know ``from which angles to teach'' or ``what kind of knowledge it needs.'' P06 noted that ``there's some information I'm not even sure I need to provide.'' P10 questioned whether the information provided overlapped with the robot's needs, potentially resulting in ``information that isn't useful.'' These accounts locate part of the teaching effort in inferring the system's information requirements before deciding what to say.

\emph{Concern about knowledge omissions.} Participants worried about leaving relevant information untaught. P14 described uncertainty about ``whether I left out some information'' when the robot did not ask questions. P07 associated omissions with simultaneously teaching and tidying, while P15 would omit information that seemed self-evident. This concern could also persist under robot initiative. Asked to explain a high Post-Task gap-monitoring rating, P18 wondered whether unasked information was unnecessary or had been overlooked, explaining that ``I couldn't tell who was responsible for the rest.'' Robot questions could reduce the need to generate teaching topics without fully clarifying responsibility for what remained untaught.

\emph{Robot-surfaced knowledge gaps.} Questions prompted participants to articulate uses they would not have volunteered. P05 had not considered explaining what different cups were used to drink. When asked what a laptop was mainly used for, P15 realized that its role in working and studying at home was information the robot needed, adding that usage and habitual location were things ``I basically wouldn't say on my own.'' Recognizing this benefit did not necessarily imply preferring robot initiative. P17 acknowledged that some questions were valuable, but considered only a few objects worth questioning and still preferred User-Led Teaching. These accounts distinguish the value of an individual question from acceptance of questioning across all objects.

\textbf{RQ2a: Situational cue availability.} \emph{Object cues during handling.} Participants described the visible object and ongoing action as resources for answering. P15 explained that ``the thing is right in my hand'' and ``I don't have to recall anything; I can say it while looking at it.'' P07 was already considering an object's function or ownership while placing it, making an immediate answer available. However, access to these cues did not eliminate all interpretive work. P06 described comparing similar cups' features and relating experimental objects to objects at home. This account qualifies the assumption that an object in view always makes its referent or personal meaning immediately clear.

\emph{Post-task context reconstruction.} Participants described recovering earlier actions and placements when answering after tidying. P04 found the robot's references clear but still needed to ``recall what you did before.'' This recovery could involve looking at the objects again. P11 had to ``think back, and I might even need to go look.'' In a corresponding session observation, P07 paused and returned to the table to inspect an object before answering. P09 linked recall difficulty to small objects placed casually, explaining that a placement might occur ``without thinking, just putting it down.'' P05, by contrast, attributed relatively easy recall to familiarity with commonly used objects that could come to mind immediately. These accounts locate post-task reconstruction in recovering specific actions and placements, with reinspection offering a way to revisit the scene before answering.

\textbf{RQ2b: Experiences of question timing.} \emph{Conversational immediacy.} P11 and P15 experienced in-task questions as natural dialogue. P11 found ``chatting while working quite pleasant'' and felt more interrupted by questions after tidying. P15 described the interaction as ``very natural, like chatting,'' while acknowledging occasional pauses of a few seconds. P15 accepted these pauses because they avoided effortful recall. Conversational flow and brief interruptions could therefore coexist within the same participant's experience.

\emph{Task-flow interruption.} Participants described questions as disrupting ongoing actions. P13 felt that the system's many detailed questions were ``a bit disruptive.'' P12 had already moved to subsequent objects while the system was still asking about the first, and would have preferred in-task questioning without these delays. Reports of latency identify a possible contributor to disruption without isolating its effect. Interruption nevertheless had different consequences for preference. P18 accepted momentary pauses to avoid lingering uncertainty about whether information was missing, explaining, ``I'd rather be interrupted a few times on the spot than keep worrying afterwards.'' This account concerns reassurance about teaching completeness, distinct from P15's emphasis on avoiding recall.

\emph{Post-task answering comfort.} Participants also valued completing the task before answering. P08 described post-task questions as the least disruptive option, and P10 preferred answering ``afterwards, in a comfortable place.'' P10 acknowledged some recall effort but felt the comfortable setting could offset it. P06 found that the overall placement rationale was clearer after tidying and that answering then required little effort. These accounts distinguish accepting some recall effort in exchange for comfort from experiencing post-task recall as relatively easy. Both could make deferred questioning attractive.

\textbf{Cross-cutting theme: Control, correction, and adoption conditions.} Participants wanted to verify what the robot had learned and regulate questioning. P08 explained that, after receiving a user's answer, the robot's spoken acknowledgement, ``Okay, I've remembered it,'' did not reveal what information it had retained: ``What did you remember? I don't know what you remembered.'' P15 requested a way to pause questions when busy. P09 wanted question timing to reflect placement habits, avoiding immediate questions for objects with fixed locations while asking about variably placed objects in context. P13 raised concerns about data handling, requesting that information be processed locally and ``not uploaded to the company's databases.'' These accounts motivate informative spoken feedback, control over questioning, and local data processing, informing the design implications in Section~\ref{sec:discussion:implications}.

\section{Discussion}
\label{sec:discussion}

Preference rankings reflected different priorities in teaching. Participants who preferred User-Led Teaching valued control over what, when, and whether to teach; those who preferred Post-Task Questioning valued task continuity; EgoAsk received the most first-place rankings, with interviewees valuing questions tied to the ongoing activity. Overall workload did not differ significantly, while teaching-specific measures indicated lower monitoring burden with robot initiative and a lower reported need for context reconstruction with in-task questions. Evidence for increased interruption was tentative. These patterns do not establish equal total effort or a direct conversion between burdens. Building on robot active learning~\cite{cakmak2010designing} and situated interaction~\cite{brown1989situated}, we examine how questions change users' responsibility for teaching, access to answering context, and opportunities to respond.

\subsection{Knowledge-Gap Monitoring}
\label{sec:discussion:monitoring}

Robot initiative can reduce the work of deciding what to teach, without necessarily clarifying the full scope of teaching. Both robot-initiated conditions elicited lower knowledge-gap monitoring burden than User-Led Teaching, supporting the expected benefit of proactive questions. Interviews suggest that concrete questions helped users identify information worth providing instead of independently anticipating the robot's needs. This benefit appeared in both questioning arrangements, although the comparison does not establish equivalent effects of their timing.

Knowing one's own routines does not imply knowing which details a robot requires. User-led teaching therefore involves both expressing personal knowledge and judging its relevance to the system, consistent with prior work on aligning robot learning with human teaching behavior~\cite{thomaz2008teachable}. A concrete question makes a particular use or habit relevant to the exchange, allowing users to explain familiar knowledge without first selecting a teaching topic. This interpretation connects our findings to research on how robot questions communicate learning needs and influence teaching~\cite{cakmak2010designing,chao2010transparent}. The benefit concerns responsibility for selecting information, beyond the effort of articulating an answer.

Responsibility for unasked information can nevertheless remain unresolved. P18 could answer individual questions but was unsure whether other information was unnecessary or had been overlooked. Prior work shows that learning feedback can help teachers assess a robot's knowledge~\cite{habibian2022learned}; P18's account raises the complementary issue of who remains responsible for information outside the exchange. This case suggests that clear individual questions need not make the overall teaching scope clear. Robot initiative may relieve users of selecting topics while leaving them uncertain about whether further teaching is expected.

\subsection{Situated Answering}
\label{sec:discussion:situated}

In-task questioning may ease answering by preserving activity context, beyond making the queried object identifiable. EgoAsk elicited a lower reported need to reconstruct the handling context than Post-Task Questioning, providing preliminary, item-level support for the expected contextual benefit. Building on the reduced topic-selection effort discussed above, this finding concerns a different requirement: users must also have access to the context needed to formulate an answer.

Object identification and activity reconstruction can require different cues. P04 understood which object the robot referred to but still needed to recall earlier actions. Collaborative reference helps establish the object under discussion~\cite{clark1986referring}, whereas explaining its use or placement may additionally depend on recovering actions and surrounding circumstances. Encoding specificity and context-dependent memory offer a plausible explanation for the role of these cues~\cite{tulving1973encoding,smith2001environmental}. Related systems situate knowledge discovery in everyday encounters~\cite{cai2025aiget} and elicit subjective rules through shared first-person interaction~\cite{teng2026eye2eye}. Our findings highlight a complementary role for first-person context: supporting users' explanations of their actions, rather than only establishing what an object is.

Dependence on immediate activity cues varies across situations. Familiar objects and the completed arrangement could make post-task answering easier, whereas casual placements could require more reconstruction. These accounts qualify the contextual benefit without implying that immediate questioning is always necessary. The quantitative evidence concerns reported reconstruction need rather than a validated composite measure, and Post-Task Questioning combined delay, batching, and moving away from the table. The study therefore does not isolate a memory mechanism or an effect of delay alone. Its more specific insight is that understanding which object is being discussed can leave the activity context needed to explain it unresolved.

\subsection{Question Timing}
\label{sec:discussion:timing}

The acceptability of a question depends on how it fits the user's ongoing activity, not simply on whether it is asked during or after the task. Evidence for greater interruption burden with EgoAsk than Post-Task Questioning was tentative, without a significant overall condition effect. Interviews nevertheless showed why participants could prefer in-task questions while acknowledging pauses: some accepted interruptions to avoid recall or uncertainty about omissions, while others valued uninterrupted activity. Related robot-curiosity research also reports preference alongside greater effort~\cite{leusmann2025curiosity}, although effort and interruption are distinct measures. Preference thus reflects reasons for accepting an exchange, rather than an absence of disruption.

As activities progress, a useful question can become misaligned with a convenient answering moment. Task boundaries influence interruption costs~\cite{adamczyk2004not,iqbal2006leveraging}. P12 described working with subsequent objects while the robot was still asking about an earlier one. This account directly identifies a mismatch between question delivery and task progress; it does not establish that answering cues had been lost. Such a mismatch may nevertheless require users to redirect attention to an earlier activity while engaged in the current one. This connects question timing to the contextual requirements discussed above: information can remain worth acquiring even when the circumstances that made it convenient to explain have changed.

The importance of this alignment also depends on the knowledge sought and the user's priorities. P09 favored contextual questions about variable placements but not repeated immediate questions about fixed locations. Other participants valued the comfort of post-task answering, even when it involved some recall. Task Matters likewise identifies task-dependent questioning sequences in human interaction~\cite{hu2025taskmatters}. These accounts suggest complementary roles for different teaching arrangements, without establishing an optimal schedule or isolating latency's contribution. Recognizing a question's value does not entail being willing to answer it at any moment.

\subsection{Design Implications}
\label{sec:discussion:implications}

\textbf{Make teaching scope legible.} Systems should make teaching scope and responsibilities explicit, helping users judge what the robot has learned and what they may still need to contribute. They could distinguish recorded information from identified but unresolved knowledge gaps, while retaining an entry point for user-led additions. At the end of a questioning session, a brief summary should describe what was recorded without implying that all relevant knowledge has been acquired. This makes the allocation of initiative between users and systems~\cite{horvitz1999principles} understandable as a division of teaching responsibilities, allowing users to decide whether to contribute further rather than infer completeness from the robot's silence.

\textbf{Preserve activity context.} Systems should support recovery of the activity context needed to answer, beyond identifying the object in question. Situated cognition emphasizes the relation between knowledge and activity~\cite{brown1989situated}, and our interviews indicate that earlier actions may still require recall after the object is understood. When users cannot answer immediately, systems could retain question-relevant action and location cues with their permission and present them when asking later. RetroSketch's video-supported retrospective reporting offers a design precedent, although its evaluation concerns VR experience rather than object teaching~\cite{potts2025retrosketch}. These cues should help users move from recognizing an object to recovering how they handled it, providing a basis for explaining their actions.

\textbf{Reassess questions before delivery.} Systems should reassess whether users can conveniently answer at delivery time, rather than ask solely on the basis of a question's value when generated. ProMemAssist defers, reassesses, or discards assistance as working-memory conditions change~\cite{pu2025promemassist}, while Task Matters identifies task-dependent questioning sequences~\cite{hu2025taskmatters}. Systems could accordingly consider the current object, answering cues, and task progress when deciding whether to ask immediately, defer for review, or drop the current question, while allowing users to volunteer information, pause, or postpone an exchange. Such scheduling aims to fit knowledge acquisition to users' activity rather than require them to return to an earlier task for a question whose convenient answering moment has passed.

\textbf{Confirm the stored interpretation.} Systems should let users verify the interpretation actually stored, beyond acknowledging receipt of an answer. Grounding requires opportunities to check mutual understanding~\cite{clark1991grounding}, and learning feedback can help teachers assess a robot's knowledge~\cite{habibian2022learned}. A confirmation should identify the associated object, recorded fact, and qualifications such as ``usually.'' When scope is unclear, systems should clarify whether an answer describes a single occasion or an enduring habit and provide an immediate correction or undo option. Such confirmation aims to expose errors in object assignment or interpretation before a misunderstood answer becomes the basis for subsequent assistance.

\textbf{Keep personal memory revisable.} Systems should support ongoing management of personal knowledge as household routines and users' willingness to share or use information change. Human-AI interaction guidelines distinguish local feedback from global control~\cite{amershi2019guidelines}; beyond verifying individual exchanges, users should be able to inspect, update, delete, and restrict the use of knowledge. Natural-language rules can express conditional preferences~\cite{li2019pumice}, and systems should preserve these conditions for later revision. They should also address participants' data-handling concerns by making storage locations and uses explicit. This makes each teaching exchange part of a revisable knowledge record rather than permanent authorization for every future context and use.

Together, the findings characterize personal object teaching as more than the delivery of clear, relevant questions. Robot initiative may leave teaching responsibility unresolved; an identifiable object may lack the activity context needed to explain it; and a useful question may arrive after a convenient answering moment has passed. These distinctions connect the comparative burden findings to the conditions under which users can explain, verify, and revise knowledge for household robot services.

\subsection{Limitations}
\label{sec:discussion:limitations}

\textbf{Study Limitations.} Our evaluation involved a single tidying session in a simulated household. It does not establish how teaching burden and preferences evolve with repeated use, changing household routines, or more time-sensitive activities. Longer-term in-home deployments should examine whether the observed benefits persist as users become familiar with the system and repeatedly teach or revise object knowledge. The evaluation also focused on subjective burden, preferences, and interview accounts, with context-reconstruction evidence limited to individual questionnaire items. We did not measure the accuracy or coverage of the acquired knowledge. Future studies should assess these outcomes alongside user effort to determine whether a more acceptable teaching experience leads to reliable knowledge and improved household assistance.

\textbf{System Limitations.} Participants reported difficulties with object recognition, speech input, and delayed questions in the current prototype. In particular, questions about a previous object could arrive after users had moved on to another. Future iterations should profile the processing pipeline and evaluate sentence-level streaming speech synthesis and, where supported by the inference backend, reuse of cached context to reduce waiting, following technical approaches explored in AURA~\cite{lu2026aura}. Faster processing alone does not ensure appropriate timing. ProMemAssist provides a complementary approach by deferring, reassessing, or discarding assistance as the user's modeled cognitive state changes~\cite{pu2025promemassist}. For EgoAsk, this suggests rechecking the current object and activity before delivering a delayed question. Future evaluations should vary processing delay and delivery timing separately, measuring task disruption alongside question relevance and knowledge accuracy to establish whether these changes improve teaching without compromising reliability.

\section{Conclusion}
\label{sec:conclusion}

We presented EgoAsk, a smart-glasses system that enables household robots to acquire personalized object knowledge through questions grounded in the user's first-person view and ongoing activity. A within-subjects study with 18 participants showed that robot-initiated questioning reduced perceived knowledge-gap monitoring burden relative to user-led teaching, while EgoAsk elicited a lower reported need to reconstruct handling context than post-task questioning. Overall workload did not differ significantly across conditions. Interviews highlighted differing priorities: some participants valued answering while relevant cues remained available, whereas others preferred uninterrupted activity and deferred exchanges. These findings suggest that integrating robot teaching into everyday life requires coordinating what the robot needs to learn with when users can explain it. Proactive questions can guide teaching, but their usefulness also depends on available context, task progress, and user control. Future work should investigate adaptive question timing and longer-term use in homes, evaluating the accuracy and coverage of acquired knowledge alongside teaching burden and its value for subsequent household assistance.

\bibliographystyle{ACM-Reference-Format}
\bibliography{refs}

@inproceedings{kapelyukh2022myhouse,
  author         = {Kapelyukh, Ivan and Johns, Edward},
  title          = {My House, My Rules: Learning Tidying Preferences with Graph Neural Networks},
  booktitle      = {Proceedings of the 5th Conference on Robot Learning (CoRL)},
  series         = {Proceedings of Machine Learning Research},
  volume         = {164},
  pages          = {740--749},
  year           = {2022},
  url            = {https://proceedings.mlr.press/v164/kapelyukh22a.html}
}

@inproceedings{grauman2022ego4d,
  author         = {Grauman, Kristen and Westbury, Andrew and Byrne, Eugene and Chavis, Zachary and Furnari, Antonino and Girdhar, Rohit and Hamburger, Jackson and Jiang, Hao and Liu, Miao and Liu, Xingyu and Martin, Miguel and Nagarajan, Tushar and Radosavovic, Ilija and Ramakrishnan, Santhosh Kumar and Ryan, Fiona and Sharma, Jayant and Wray, Michael and Xu, Mengmeng and Xu, Eric Zhongcong and Zhao, Chen and Bansal, Siddhant and Batra, Dhruv and Cartillier, Vincent and Crane, Sean and Do, Tien and Doulaty, Morrie and Erapalli, Akshay and Feichtenhofer, Christoph and Fragomeni, Adriano and Fu, Qichen and Gebreselasie, Abrham and Gonz{\'a}lez, Cristina and Hillis, James and Huang, Xuhua and Huang, Yifei and Jia, Wenqi and Khoo, Weslie and Kol{\'a}{\v{r}}, J{\'a}chym and Kottur, Satwik and Kumar, Anurag and Landini, Federico and Li, Chao and Li, Yanghao and Li, Zhenqiang and Mangalam, Karttikeya and Modhugu, Raghava and Munro, Jonathan and Murrell, Tullie and Nishiyasu, Takumi and Price, Will and Ruiz, Paola and Ramazanova, Merey and Sari, Leda and Somasundaram, Kiran and Southerland, Audrey and Sugano, Yusuke and Tao, Ruijie and Vo, Minh and Wang, Yuchen and Wu, Xindi and Yagi, Takuma and Zhao, Ziwei and Zhu, Yunyi and Arbel{\'a}ez, Pablo and Crandall, David and Damen, Dima and Farinella, Giovanni Maria and Fuegen, Christian and Ghanem, Bernard and Ithapu, Vamsi Krishna and Jawahar, C. V. and Joo, Hanbyul and Kitani, Kris and Li, Haizhou and Newcombe, Richard and Oliva, Aude and Park, Hyun Soo and Rehg, James M. and Sato, Yoichi and Shi, Jianbo and Shou, Mike Zheng and Torralba, Antonio and Torresani, Lorenzo and Yan, Mingfei and Malik, Jitendra},
  title          = {{Ego4D}: Around the World in 3,000 Hours of Egocentric Video},
  booktitle      = {Proceedings of the IEEE/CVF Conference on Computer Vision and Pattern Recognition (CVPR)},
  pages          = {18995--19012},
  year           = {2022},
  doi            = {10.1109/CVPR52688.2022.01842}
}

@inproceedings{leusmann2025curiosity,
  author         = {Leusmann, Jan and Belardinelli, Anna and Haliburton, Luke and Hasler, Stephan and Schmidt, Albrecht and Mayer, Sven and Gienger, Michael and Wang, Chao},
  title          = {Investigating {LLM}-Driven Curiosity in Human-Robot Interaction},
  booktitle      = {Proceedings of the 2025 CHI Conference on Human Factors in Computing Systems},
  year           = {2025},
  publisher      = {Association for Computing Machinery},
  numpages       = {16},
  doi            = {10.1145/3706598.3713923},
  url            = {https://doi.org/10.1145/3706598.3713923},
  articleno      = {599},
  pages          = {1--16}
}

@misc{hu2025taskmatters,
  author         = {Hu, Yuanda and Hou, Jiani and Zhang, Junyu and Ge, Yate and Sun, Xiaohua and Guo, Weiwei},
  title          = {Task Matters: Investigating Human Questioning Behavior in Different Household Service for Learning by Asking Robots},
  year           = {2025},
  eprint         = {2504.13916},
  archivePrefix  = {arXiv},
  url            = {https://arxiv.org/abs/2504.13916},
  primaryClass   = {cs.HC}
}

@inproceedings{ramachandruni2025parsec,
  author         = {Ramachandruni, Kartik and Chernova, Sonia},
  title          = {Personalized Robotic Object Rearrangement from Scene Context},
  booktitle      = {IEEE International Conference on Robot and Human Interactive Communication (RO-MAN)},
  year           = {2025},
  url            = {https://doi.org/10.1109/RO-MAN63969.2025.11217903},
  pages          = {259--266},
  doi            = {10.1109/RO-MAN63969.2025.11217903}
}

@misc{newman2024degustabot,
  author         = {Newman, Benjamin A. and Gupta, Pranay and Kitani, Kris and Bisk, Yonatan and Admoni, Henny and Paxton, Chris},
  title          = {{DegustaBot}: Zero-Shot Visual Preference Estimation for Personalized Multi-Object Rearrangement},
  year           = {2024},
  url            = {https://arxiv.org/abs/2407.08876},
  primaryClass   = {cs.CV},
  eprint         = {2407.08876},
  archivePrefix  = {arXiv}
}

@inproceedings{lee2026bringmycup,
  title          = {Bring My Cup! Personalizing Vision-Language-Action Models with Visual Attentive Prompting},
  author         = {Sangoh Lee and Sangwoo Mo and Wook-Shin Han},
  year           = {2026},
  url            = {https://vap-project.github.io/},
  booktitle      = {Proceedings of the International Conference on Machine Learning (ICML)},
  eprint         = {2512.20014},
  archivePrefix  = {arXiv},
  primaryClass   = {cs.RO}
}

@article{ikeuchi2024semantic,
  author  = {Ikeuchi, Katsushi and Wake, Naoki and Sasabuchi, Kazuhiro and Takamatsu, Jun},
  title   = {Semantic Constraints to Represent Common Sense Required in Household Actions for Multimodal Learning-from-Observation Robot},
  journal = {The International Journal of Robotics Research},
  volume  = {43},
  number  = {2},
  pages   = {134--170},
  year    = {2024},
  doi     = {10.1177/02783649231212929}
}

@article{wu2023tidybot,
  author  = {Wu, Jimmy and Antonova, Rika and Kan, Adam and Lepert, Marion and Zeng, Andy and Song, Shuran and Bohg, Jeannette and Rusinkiewicz, Szymon and Funkhouser, Thomas},
  title   = {{TidyBot}: Personalized Robot Assistance with Large Language Models},
  journal = {Autonomous Robots},
  volume  = {47},
  number  = {8},
  pages   = {1087--1102},
  year    = {2023},
  doi     = {10.1007/s10514-023-10139-z}
}

@inproceedings{barsellotti2024personalized,
  author    = {Barsellotti, Luca and Bigazzi, Roberto and Cornia, Marcella and Baraldi, Lorenzo and Cucchiara, Rita},
  title     = {Personalized Instance-Based Navigation Toward User-Specific Objects in Realistic Environments},
  booktitle = {Advances in Neural Information Processing Systems (NeurIPS)},
  volume    = {37},
  pages     = {11228--11250},
  year      = {2024},
  doi       = {10.52202/079017-0358}
}

@inproceedings{dai2024think,
  author    = {Dai, Yinpei and Peng, Run and Li, Sikai and Chai, Joyce},
  title     = {Think, Act, and Ask: Open-World Interactive Personalized Robot Navigation},
  booktitle = {Proceedings of the IEEE International Conference on Robotics and Automation (ICRA)},
  pages     = {3296--3303},
  year      = {2024},
  doi       = {10.1109/ICRA57147.2024.10610178}
}

@misc{liang2026pahf,
  author        = {Liang, Kaiqu and Kruk, Julia and Qian, Shengyi and Yang, Xianjun and Bi, Shengjie and Yao, Yuanshun and Nie, Shaoliang and Zhang, Mingyang and Liu, Lijuan and Fisac, Jaime Fern{\'a}ndez and Zhou, Shuyan and Hosseini, Saghar},
  title         = {Learning Personalized Agents from Human Feedback},
  year          = {2026},
  eprint        = {2602.16173},
  archiveprefix = {arXiv},
  primaryclass  = {cs.AI}
}

@article{thomaz2008teachable,
  author  = {Thomaz, Andrea L. and Breazeal, Cynthia},
  title   = {Teachable Robots: Understanding Human Teaching Behavior to Build More Effective Robot Learners},
  journal = {Artificial Intelligence},
  volume  = {172},
  number  = {6--7},
  pages   = {716--737},
  year    = {2008},
  doi     = {10.1016/j.artint.2007.09.009}
}

@article{cakmak2010designing,
  author  = {Cakmak, Maya and Chao, Crystal and Thomaz, Andrea L.},
  title   = {Designing Interactions for Robot Active Learners},
  journal = {{IEEE} Transactions on Autonomous Mental Development},
  volume  = {2},
  number  = {2},
  pages   = {108--118},
  year    = {2010},
  doi     = {10.1109/TAMD.2010.2051030}
}

@article{krishna2022socially,
  author  = {Krishna, Ranjay and Lee, Donsuk and Fei-Fei, Li and Bernstein, Michael S.},
  title   = {Socially situated artificial intelligence enables learning from human interaction},
  journal = {Proceedings of the National Academy of Sciences},
  volume  = {119},
  number  = {39},
  pages   = {e2115730119},
  year    = {2022},
  doi     = {10.1073/pnas.2115730119}
}

@inproceedings{horvitz1999principles,
  author    = {Horvitz, Eric},
  title     = {Principles of Mixed-Initiative User Interfaces},
  booktitle = {Proceedings of the SIGCHI Conference on Human Factors in Computing Systems (CHI '99)},
  year      = {1999},
  pages     = {159--166},
  publisher = {ACM Press},
  doi       = {10.1145/302979.303030}
}

@inproceedings{chao2010transparent,
  author    = {Chao, Crystal and Cakmak, Maya and Thomaz, Andrea L.},
  title     = {Transparent Active Learning for Robots},
  booktitle = {Proceedings of the 5th ACM/IEEE International Conference on Human-Robot Interaction (HRI '10)},
  year      = {2010},
  pages     = {317--324},
  publisher = {ACM/IEEE},
  doi       = {10.1145/1734454.1734562}
}

@inproceedings{cai2025aiget,
  author    = {Cai, Runze and Janaka, Nuwan and Kim, Hyeongcheol and Chen, Yang and Zhao, Shengdong and Huang, Yun and Hsu, David},
  title     = {{AiGet}: Transforming Everyday Moments into Hidden Knowledge Discovery with {AI} Assistance on Smart Glasses},
  booktitle = {Proceedings of the {CHI} Conference on Human Factors in Computing Systems},
  series    = {{CHI} '25},
  articleno = {631},
  pages     = {1--26},
  year      = {2025},
  doi       = {10.1145/3706598.3713953},
  publisher = {{ACM}}
}

@inproceedings{cakmak2012designing,
  author    = {Cakmak, Maya and Thomaz, Andrea L.},
  title     = {Designing Robot Learners That Ask Good Questions},
  booktitle = {Proceedings of the 7th ACM/IEEE International Conference on Human-Robot Interaction (HRI)},
  pages     = {17--24},
  year      = {2012},
  doi       = {10.1145/2157689.2157693}
}

@inproceedings{li2019pumice,
  author    = {Li, Toby Jia-Jun and Radensky, Marissa and Jia, Justin and Singarajah, Kirielle and Mitchell, Tom M. and Myers, Brad A.},
  title     = {{PUMICE}: A Multi-Modal Agent That Learns Concepts and Conditionals from Natural Language and Demonstrations},
  booktitle = {Proceedings of the 32nd Annual ACM Symposium on User Interface Software and Technology (UIST)},
  pages     = {577--589},
  year      = {2019},
  doi       = {10.1145/3332165.3347899}
}

@inproceedings{fanello2013icubworld,
  author    = {Fanello, Sean R. and Ciliberto, Carlo and Santoro, Matteo and Natale, Lorenzo and Metta, Giorgio and Rosasco, Lorenzo and Odone, Francesca},
  title     = {{iCub} World: Friendly Robots Help Building Good Vision Data-Sets},
  booktitle = {Proceedings of the IEEE Conference on Computer Vision and Pattern Recognition Workshops (CVPRW)},
  pages     = {700--705},
  year      = {2013},
  doi       = {10.1109/CVPRW.2013.106}
}

@inproceedings{pasquale2015teaching,
  author    = {Pasquale, Giulia and Ciliberto, Carlo and Odone, Francesca and Rosasco, Lorenzo and Natale, Lorenzo},
  title     = {Teaching {iCub} to Recognize Objects Using Deep Convolutional Neural Networks},
  booktitle = {Proceedings of the 4th Workshop on Machine Learning for Interactive Systems (MLIS) at ICML},
  series    = {JMLR Workshop and Conference Proceedings},
  volume    = {43},
  pages     = {21--25},
  year      = {2015}
}

@inproceedings{maiettini2022handheld,
  author        = {Maiettini, Elisa and Maracani, Andrea and Camoriano, Raffaello and Pasquale, Giulia and Tikhanoff, Vadim and Rosasco, Lorenzo and Natale, Lorenzo},
  title         = {From Handheld to Unconstrained Object Detection: A Weakly-Supervised On-Line Learning Approach},
  booktitle     = {Proceedings of the IEEE International Conference on Robot and Human Interactive Communication (RO-MAN)},
  pages         = {942--949},
  year          = {2022},
  doi           = {10.1109/RO-MAN53752.2022.9900780},
  eprint        = {2012.14345},
  archiveprefix = {arXiv}
}

@inproceedings{ayub2020tellme,
  author    = {Ayub, Ali and Wagner, Alan R.},
  title     = {Tell Me What This Is: Few-Shot Incremental Object Learning by a Robot},
  booktitle = {Proceedings of the IEEE/RSJ International Conference on Intelligent Robots and Systems (IROS)},
  pages     = {8344--8350},
  year      = {2020},
  doi       = {10.1109/IROS45743.2020.9341140}
}

@inproceedings{kareer2025egomimic,
  author        = {Kareer, Simar and Patel, Dhruv and Punamiya, Ryan and Mathur, Pranay and Cheng, Shuo and Wang, Chen and Hoffman, Judy and Xu, Danfei},
  title         = {{EgoMimic}: Scaling Imitation Learning via Egocentric Video},
  booktitle     = {2025 IEEE International Conference on Robotics and Automation (ICRA)},
  year          = {2025},
  eprint        = {2410.24221},
  archiveprefix = {arXiv},
  pages         = {13226--13233},
  doi           = {10.1109/ICRA55743.2025.11127989}
}

@misc{liu2025egozero,
  author        = {Liu, Vincent and Adeniji, Ademi and Zhan, Haotian and Haldar, Siddhant and Bhirangi, Raunaq and Abbeel, Pieter and Pinto, Lerrel},
  title         = {{EgoZero}: Robot Learning from Smart Glasses},
  year          = {2025},
  eprint        = {2505.20290},
  archiveprefix = {arXiv},
  primaryclass  = {cs.RO}
}

@inproceedings{suzuki2022augmented,
  author    = {Suzuki, Ryo and Karim, Adnan and Xia, Tian and Hedayati, Hooman and Marquardt, Nicolai},
  title     = {Augmented Reality and Robotics: A Survey and Taxonomy for {AR}-Enhanced Human-Robot Interaction and Robotic Interfaces},
  booktitle = {Proceedings of the 2022 CHI Conference on Human Factors in Computing Systems (CHI)},
  articleno = {553},
  pages     = {1--33},
  year      = {2022},
  doi       = {10.1145/3491102.3517719}
}

@misc{jaykumar2024iteach,
  author        = {Jaykumar P, Jishnu and Salvato, Cole and Bomnale, Vinaya and Wang, Jikai and Xiang, Yu},
  title         = {{iTeach}: Interactive Teaching for Robot Perception Using Mixed Reality},
  year          = {2024},
  eprint        = {2410.09072v1},
  url           = {https://arxiv.org/abs/2410.09072v1},
  archiveprefix = {arXiv},
  primaryclass  = {cs.RO}
}

@article{belardinelli2025train,
  author  = {Belardinelli, Anna and Wang, Chao and Tanneberg, Daniel and Hasler, Stephan and Gienger, Michael},
  title   = {Train Your Robot in {AR}: Insights and Challenges for Humans and Robots in Continual Teaching and Learning},
  journal = {Frontiers in Robotics and AI},
  volume  = {12},
  pages   = {1605652},
  year    = {2025},
  doi     = {10.3389/frobt.2025.1605652}
}

@article{ayub2024human,
  author    = {Ayub, Ali and De Francesco, Zachary and Mehta, Jainish and Agha, Khaled Yaakoub and Holthaus, Patrick and Nehaniv, Chrystopher L. and Dautenhahn, Kerstin},
  title     = {A Human-Centered View of Continual Learning: Understanding Interactions, Teaching Patterns, and Perceptions of Human Users Toward a Continual Learning Robot in Repeated Interactions},
  journal   = {ACM Transactions on Human-Robot Interaction},
  volume    = {13},
  number    = {4},
  articleno = {52},
  pages     = {1--39},
  year      = {2024},
  doi       = {10.1145/3659110}
}

@inproceedings{ayub2024interactive,
  author        = {Ayub, Ali and Nehaniv, Chrystopher L. and Dautenhahn, Kerstin},
  title         = {Interactive Continual Learning Architecture for Long-Term Personalization of Home Service Robots},
  booktitle     = {2024 IEEE International Conference on Robotics and Automation (ICRA)},
  year          = {2024},
  eprint        = {2403.03462},
  archiveprefix = {arXiv},
  pages         = {11289--11296},
  doi           = {10.1109/ICRA57147.2024.10611386}
}

@inproceedings{wang2020see,
  author    = {Wang, Yeping and Ajaykumar, Gopika and Huang, Chien-Ming},
  title     = {See What {I} See: Enabling User-Centric Robotic Assistance Using First-Person Demonstrations},
  booktitle = {Proceedings of the 2020 ACM/IEEE International Conference on Human-Robot Interaction (HRI)},
  pages     = {639--648},
  year      = {2020},
  doi       = {10.1145/3319502.3374820}
}

@inproceedings{sadigh2017active,
  author    = {Sadigh, Dorsa and Dragan, Anca D. and Sastry, Shankar and Seshia, Sanjit A.},
  title     = {Active Preference-Based Learning of Reward Functions},
  booktitle = {Proceedings of Robotics: Science and Systems (RSS)},
  year      = {2017},
  doi       = {10.15607/RSS.2017.XIII.053}
}

@incollection{hart1988tlx,
  author    = {Hart, Sandra G. and Staveland, Lowell E.},
  title     = {Development of {NASA-TLX} ({Task Load Index}): Results of Empirical and Theoretical Research},
  booktitle = {Human Mental Workload},
  editor    = {Hancock, Peter A. and Meshkati, Najmedin},
  series    = {Advances in Psychology},
  volume    = {52},
  publisher = {North-Holland},
  pages     = {139--183},
  year      = {1988},
  doi       = {10.1016/S0166-4115(08)62386-9}
}

@article{tulving1973encoding,
  author  = {Tulving, Endel and Thomson, Donald M.},
  title   = {Encoding Specificity and Retrieval Processes in Episodic Memory},
  journal = {Psychological Review},
  volume  = {80},
  number  = {5},
  year    = {1973},
  pages   = {352--373},
  doi     = {10.1037/h0020071}
}

@article{smith2001environmental,
  author  = {Smith, Steven M. and Vela, Edward},
  title   = {Environmental Context-Dependent Memory: A Review and Meta-Analysis},
  journal = {Psychonomic Bulletin \& Review},
  volume  = {8},
  number  = {2},
  year    = {2001},
  pages   = {203--220},
  doi     = {10.3758/BF03196157}
}

@article{clark1986referring,
  author  = {Clark, Herbert H. and Wilkes-Gibbs, Deanna},
  title   = {Referring as a Collaborative Process},
  journal = {Cognition},
  volume  = {22},
  number  = {1},
  year    = {1986},
  pages   = {1--39},
  doi     = {10.1016/0010-0277(86)90010-7}
}

@incollection{clark1991grounding,
  author    = {Clark, Herbert H. and Brennan, Susan E.},
  title     = {Grounding in Communication},
  booktitle = {Perspectives on Socially Shared Cognition},
  editor    = {Resnick, Lauren B. and Levine, John M. and Teasley, Stephanie D.},
  year      = {1991},
  pages     = {127--149},
  publisher = {American Psychological Association},
  address   = {Washington, DC},
  doi       = {10.1037/10096-006}
}

@inproceedings{adamczyk2004not,
  author    = {Adamczyk, Piotr D. and Bailey, Brian P.},
  title     = {If Not Now, When? The Effects of Interruption at Different Moments Within Task Execution},
  booktitle = {Proceedings of the CHI Conference on Human Factors in Computing Systems},
  pages     = {271--278},
  year      = {2004},
  doi       = {10.1145/985692.985727}
}

@inproceedings{iqbal2006leveraging,
  author    = {Iqbal, Shamsi T. and Bailey, Brian P.},
  title     = {Leveraging Characteristics of Task Structure to Predict the Cost of Interruption},
  booktitle = {Proceedings of the SIGCHI Conference on Human Factors in Computing Systems (CHI '06)},
  year      = {2006},
  pages     = {741--750},
  publisher = {ACM},
  doi       = {10.1145/1124772.1124882}
}

@article{bailey2008workload,
  author    = {Bailey, Brian P. and Iqbal, Shamsi T.},
  title     = {Understanding Changes in Mental Workload during Execution of Goal-Directed Tasks and Its Application for Interruption Management},
  journal   = {ACM Transactions on Computer-Human Interaction},
  volume    = {14},
  number    = {4},
  articleno = {21},
  year      = {2008},
  doi       = {10.1145/1314683.1314689},
  numpages  = {28}
}

@article{braun2006thematic,
  author  = {Braun, Virginia and Clarke, Victoria},
  title   = {Using Thematic Analysis in Psychology},
  journal = {Qualitative Research in Psychology},
  volume  = {3},
  number  = {2},
  pages   = {77--101},
  year    = {2006},
  doi     = {10.1191/1478088706qp063oa}
}

@article{cohen1960coefficient,
  author  = {Cohen, Jacob},
  title   = {A Coefficient of Agreement for Nominal Scales},
  journal = {Educational and Psychological Measurement},
  volume  = {20},
  number  = {1},
  pages   = {37--46},
  year    = {1960},
  doi     = {10.1177/001316446002000104}
}

@article{mcdonald2019reliability,
  author    = {McDonald, Nora and Schoenebeck, Sarita and Forte, Andrea},
  title     = {Reliability and Inter-rater Reliability in Qualitative Research: Norms and Guidelines for {CSCW} and {HCI} Practice},
  journal   = {Proceedings of the ACM on Human-Computer Interaction},
  volume    = {3},
  number    = {{CSCW}},
  articleno = {72},
  pages     = {1--23},
  year      = {2019},
  doi       = {10.1145/3359174}
}

@article{hashimoto2026actowl,
  author  = {Hashimoto, Saki and Hasegawa, Shoichi and Ishikawa, Tomochika and
             Taniguchi, Akira and Hagiwara, Yoshinobu and El Hafi, Lotfi and
             Taniguchi, Tadahiro},
  title   = {Toward Ownership Understanding of Objects: Active Question
             Generation with Large Language Model and Probabilistic
             Generative Model},
  journal = {Artificial Life and Robotics},
  volume  = {31},
  number  = {2},
  pages   = {389--404},
  year    = {2026},
  doi     = {10.1007/s10015-026-01116-7}
}

@inproceedings{tan2019mine,
  author    = {Tan, Zhi-Xuan and Brawer, Jake and Scassellati, Brian},
  title     = {That's Mine! Learning Ownership Relations and Norms for Robots},
  booktitle = {Proceedings of the AAAI Conference on Artificial Intelligence},
  volume    = {33},
  number    = {1},
  pages     = {8058--8065},
  year      = {2019},
  doi       = {10.1609/aaai.v33i01.33018058}
}

@inproceedings{kwon2026memento,
  author        = {Kwon, Taeyoon and Choi, Dongwook and Kim, Hyojun and Kim, Sunghwan and Moon, Seungjun and Kwak, Beong-woo and Huang, Kuan-Hao and Yeo, Jinyoung},
  title         = {Embodied Agents Meet Personalization: Investigating Challenges and Solutions Through the Lens of Memory Utilization},
  booktitle     = {The Fourteenth International Conference on Learning Representations},
  year          = {2026},
  eprint        = {2505.16348},
  archiveprefix = {arXiv},
  url           = {https://openreview.net/forum?id=E5L43l5EIu}
}

@article{yin2026hold,
  author  = {Yin, Maoliang and Hua, Changchun and Zhang, Ying and Gao, Yiyue and Pan, Wenlong},
  title   = {Learning what you hold: A personalized object registration and recognition framework via handheld interaction},
  journal = {Biomimetic Intelligence and Robotics},
  pages   = {100356},
  year    = {2026},
  doi     = {10.1016/j.birob.2026.100356}
}

@inproceedings{nakamura2022multimodal,
  author    = {Nakamura, Hitoshi and El Hafi, Lotfi and Taniguchi, Akira and
               Hagiwara, Yoshinobu and Taniguchi, Tadahiro},
  title     = {Multimodal Object Categorization with Reduced User Load through
               Human-Robot Interaction in Mixed Reality},
  booktitle = {Proceedings of the 2022 {IEEE/RSJ} International Conference on
               Intelligent Robots and Systems ({IROS})},
  pages     = {2143--2150},
  year      = {2022},
  doi       = {10.1109/IROS47612.2022.9981374}
}

@inproceedings{weber2023multiperspective,
  author    = {Weber, Daniel and Fuhl, Wolfgang and Kasneci, Enkelejda and Zell, Andreas},
  title     = {Multiperspective Teaching of Unknown Objects via Shared-gaze-based
               Multimodal Human-Robot Interaction},
  booktitle = {Proceedings of the 2023 {ACM/IEEE} International Conference on
               Human-Robot Interaction ({HRI})},
  pages     = {544--553},
  year      = {2023},
  doi       = {10.1145/3568162.3578627}
}

@inproceedings{weber2023saliency,
  author    = {Weber, Daniel and Bolz, Valentin and Zell, Andreas and Kasneci, Enkelejda},
  title     = {Leveraging Saliency-Aware Gaze Heatmaps for Multiperspective Teaching
               of Unknown Objects},
  booktitle = {Proceedings of the 2023 {IEEE/RSJ} International Conference on
               Intelligent Robots and Systems ({IROS})},
  year      = {2023},
  pages     = {7846--7853},
  doi       = {10.1109/IROS55552.2023.10342312}
}

@inproceedings{rosen2020bidirectional,
  author    = {Rosen, Eric and Whitney, David and Fishman, Michael and Ullman, Daniel
               and Tellex, Stefanie},
  title     = {Mixed Reality as a Bidirectional Communication Interface for
               Human-Robot Interaction},
  booktitle = {Proceedings of the {IEEE/RSJ} International Conference on Intelligent
               Robots and Systems ({IROS})},
  pages     = {11431--11438},
  year      = {2020},
  doi       = {10.1109/IROS45743.2020.9340822}
}

@inproceedings{dogan2022asking,
  author    = {Do{\u{g}}an, Fethiye Irmak and Torre, Ilaria and Leite, Iolanda},
  title     = {Asking Follow-Up Clarifications to Resolve Ambiguities in
               Human-Robot Conversation},
  booktitle = {Proceedings of the {ACM/IEEE} International Conference on
               Human-Robot Interaction ({HRI})},
  pages     = {461--469},
  year      = {2022},
  doi       = {10.1109/HRI53351.2022.9889368}
}

@inproceedings{kang2023prograsp,
  author    = {Kang, Gi-Cheon and Kim, Junghyun and Kim, Jaein and Zhang, Byoung-Tak},
  title     = {{PROGrasp}: Pragmatic Human-Robot Communication for Object Grasping},
  booktitle = {{IEEE} International Conference on Robotics and Automation ({ICRA})},
  pages     = {3304--3310},
  year      = {2024},
  doi       = {10.1109/ICRA57147.2024.10610543}
}

@inproceedings{wang2024apricot,
  author    = {Wang, Huaxiaoyue and Chin, Nathaniel and Gonzalez-Pumariega, Gonzalo
               and Sun, Xiangwan and Sunkara, Neha and Pace, Maximus Adrian and
               Bohg, Jeannette and Choudhury, Sanjiban},
  title     = {{APRICOT}: Active Preference Learning and Constraint-Aware Task
               Planning with {LLMs}},
  booktitle = {Proceedings of the 8th Conference on Robot Learning},
  series    = {Proceedings of Machine Learning Research},
  volume    = {270},
  pages     = {1590--1642},
  year      = {2025},
  url       = {https://proceedings.mlr.press/v270/wang25e.html}
}

@article{brown1989situated,
  author  = {Brown, John Seely and Collins, Allan and Duguid, Paul},
  title   = {Situated Cognition and the Culture of Learning},
  journal = {Educational Researcher},
  volume  = {18},
  number  = {1},
  pages   = {32--42},
  year    = {1989},
  doi     = {10.3102/0013189X018001032}
}

@inproceedings{wang2025yoloe,
  author        = {Wang, Ao and Liu, Lihao and Chen, Hui and Lin, Zijia and Han, Jungong and Ding, Guiguang},
  title         = {{YOLOE}: Real-Time Seeing Anything},
  booktitle     = {Proceedings of the IEEE/CVF International Conference on Computer Vision (ICCV)},
  year          = {2025},
  eprint        = {2503.07465},
  archiveprefix = {arXiv},
  pages         = {24591--24602},
  url           = {https://openaccess.thecvf.com/content/ICCV2025/html/Wang_YOLOE_Real-Time_Seeing_Anything_ICCV_2025_paper.html}
}

@inproceedings{zhang2022bytetrack,
  author        = {Zhang, Yifu and Sun, Peize and Jiang, Yi and Yu, Dongdong and Weng, Fucheng and Yuan, Zehuan and Luo, Ping and Liu, Wenyu and Wang, Xinggang},
  title         = {{ByteTrack}: Multi-Object Tracking by Associating Every Detection Box},
  booktitle     = {Computer Vision -- ECCV 2022},
  year          = {2022},
  eprint        = {2110.06864},
  archiveprefix = {arXiv},
  pages         = {1--21},
  series        = {Lecture Notes in Computer Science},
  volume        = {13682},
  publisher     = {Springer Nature Switzerland},
  doi           = {10.1007/978-3-031-20047-2_1}
}

@article{oquab2023dinov2,
  author        = {Oquab, Maxime and Darcet, Timoth{\'e}e and Moutakanni, Th{\'e}o and Vo, Huy and Szafraniec, Marc and Khalidov, Vasil and Fernandez, Pierre and Haziza, Daniel and Massa, Francisco and El-Nouby, Alaaeldin and Assran, Mahmoud and Ballas, Nicolas and Galuba, Wojciech and Howes, Russell and Huang, Po-Yao and Li, Shang-Wen and Misra, Ishan and Rabbat, Michael and Sharma, Vasu and Synnaeve, Gabriel and Xu, Hu and J{\'e}gou, Herv{\'e} and Mairal, Julien and Labatut, Patrick and Joulin, Armand and Bojanowski, Piotr},
  title         = {{DINOv2}: Learning Robust Visual Features without Supervision},
  journal       = {Transactions on Machine Learning Research},
  year          = {2024},
  eprint        = {2304.07193},
  archiveprefix = {arXiv},
  url           = {https://openreview.net/forum?id=a68SUt6zFt}
}

@inproceedings{li2025satori,
  title     = {{Satori}: Towards Proactive {AR} Assistant with Belief-Desire-Intention User Modeling},
  author    = {Chenyi Li and Guande Wu and Gromit Yeuk-Yin Chan and Dishita G Turakhia and Sonia Castelo Quispe and Dong Li and Leslie Welch and Claudio Silva and Jing Qian},
  year      = {2025},
  doi       = {10.1145/3706598.3714188},
  url       = {https://doi.org/10.1145/3706598.3714188},
  booktitle = {Proceedings of the 2025 CHI Conference on Human Factors in Computing Systems},
  pages     = {1--24},
  articleno = {1229}
}

@article{damen2022rescaling,
  author  = {Damen, Dima and Doughty, Hazel and Farinella, Giovanni Maria and Furnari, Antonino and Kazakos, Evangelos and Ma, Jian and Moltisanti, Davide and Munro, Jonathan and Perrett, Toby and Price, Will and Wray, Michael},
  title   = {Rescaling Egocentric Vision: Collection, Pipeline and Challenges for {EPIC-KITCHENS-100}},
  journal = {International Journal of Computer Vision},
  volume  = {130},
  pages   = {33--55},
  year    = {2022},
  doi     = {10.1007/s11263-021-01531-2},
  url     = {https://epic-kitchens.github.io/2022},
  number  = {1}
}

@inproceedings{lin2022egovlp,
  author    = {Lin, Kevin Qinghong and Wang, Jinpeng and Soldan, Mattia and Wray, Michael and Yan, Rui and Xu, Eric Z. and Gao, Difei and Tu, Rong-Cheng and Zhao, Wenzhe and Kong, Weijie and Cai, Chengfei and Wang, Hongfa and Damen, Dima and Ghanem, Bernard and Liu, Wei and Shou, Mike Zheng},
  title     = {Egocentric Video-Language Pretraining},
  booktitle = {Advances in Neural Information Processing Systems},
  volume    = {35},
  year      = {2022},
  doi       = {10.52202/068431-0550},
  url       = {https://proceedings.neurips.cc/paper_files/paper/2022/hash/31fb284a0aaaad837d2930a610cd5e50-Abstract-Conference.html},
  pages     = {7575--7586}
}

@inproceedings{zhao2023lavila,
  author    = {Zhao, Yue and Misra, Ishan and Kr\"ahenb\"uhl, Philipp and Girdhar, Rohit},
  title     = {Learning Video Representations From Large Language Models},
  booktitle = {Proceedings of the IEEE/CVF Conference on Computer Vision and Pattern Recognition (CVPR)},
  pages     = {6586--6597},
  year      = {2023},
  url       = {https://openaccess.thecvf.com/content/CVPR2023/html/Zhao_Learning_Video_Representations_From_Large_Language_Models_CVPR_2023_paper.html}
}

@incollection{guan2023robot,
  author    = {Guan, Junhui and Liu, Zheng and Zhao, Xiaodong and Cheng, Shiwei},
  title     = {Robot Aided Intelligent Kitchen Assistance System Using Eye Tracking Based on Object Recognition},
  booktitle = {Design Studies and Intelligence Engineering},
  publisher = {IOS Press},
  year      = {2023},
  pages     = {587--596},
  doi       = {10.3233/FAIA220752},
  url       = {https://ebooks.iospress.nl/pdf/doi/10.3233/FAIA220752}
}

@inproceedings{brown2020language,
  author    = {Brown, Tom and Mann, Benjamin and Ryder, Nick and Subbiah, Melanie and Kaplan, Jared D and Dhariwal, Prafulla and Neelakantan, Arvind and Shyam, Pranav and Sastry, Girish and Askell, Amanda and Agarwal, Sandhini and Herbert-Voss, Ariel and Krueger, Gretchen and Henighan, Tom and Child, Rewon and Ramesh, Aditya and Ziegler, Daniel and Wu, Jeffrey and Winter, Clemens and Hesse, Chris and Chen, Mark and Sigler, Eric and Litwin, Mateusz and Gray, Scott and Chess, Benjamin and Clark, Jack and Berner, Christopher and McCandlish, Sam and Radford, Alec and Sutskever, Ilya and Amodei, Dario},
  title     = {Language Models are Few-Shot Learners},
  booktitle = {Advances in Neural Information Processing Systems},
  editor    = {H. Larochelle and M. Ranzato and R. Hadsell and M.F. Balcan and H. Lin},
  volume    = {33},
  pages     = {1877--1901},
  publisher = {Curran Associates, Inc.},
  year      = {2020},
  url       = {https://proceedings.neurips.cc/paper_files/paper/2020/file/1457c0d6bfcb4967418bfb8ac142f64a-Paper.pdf}
}

@inproceedings{srivastava2022behavior,
  author    = {Srivastava, Sanjana and Li, Chengshu and Lingelbach, Michael and Mart{\'i}n-Mart{\'i}n, Roberto and Xia, Fei and Vainio, Kent Elliott and Lian, Zheng and Gokmen, Cem and Buch, Shyamal and Liu, Karen and Savarese, Silvio and Gweon, Hyowon and Wu, Jiajun and Fei-Fei, Li},
  title     = {{BEHAVIOR}: Benchmark for Everyday Household Activities in Virtual, Interactive, and Ecological Environments},
  booktitle = {Proceedings of the 5th Conference on Robot Learning},
  volume    = {164},
  series    = {Proceedings of Machine Learning Research},
  pages     = {477--490},
  publisher = {PMLR},
  year      = {2022},
  url       = {https://proceedings.mlr.press/v164/srivastava22a.html}
}

@inproceedings{li2023behavior1k,
  author    = {Li, Chengshu and Zhang, Ruohan and Wong, Josiah and Gokmen, Cem and Srivastava, Sanjana and Mart{\'i}n-Mart{\'i}n, Roberto and Wang, Chen and Levine, Gabrael and Lingelbach, Michael and Sun, Jiankai and Anvari, Mona and Hwang, Minjune and Sharma, Manasi and Aydin, Arman and Bansal, Dhruva and Hunter, Samuel and Kim, Kyu-Young and Lou, Alan and Matthews, Caleb R and Villa-Renteria, Ivan and Tang, Jerry Huayang and Tang, Claire and Xia, Fei and Savarese, Silvio and Gweon, Hyowon and Liu, Karen and Wu, Jiajun and Fei-Fei, Li},
  title     = {{BEHAVIOR-1K}: A Benchmark for Embodied {AI} with 1,000 Everyday Activities and Realistic Simulation},
  booktitle = {Proceedings of The 6th Conference on Robot Learning},
  volume    = {205},
  series    = {Proceedings of Machine Learning Research},
  pages     = {80--93},
  publisher = {PMLR},
  year      = {2023},
  url       = {https://proceedings.mlr.press/v205/li23a.html}
}

@misc{behaviorKnowledgeBase,
  key          = {BEHAVIOR},
  title        = {{BEHAVIOR-1K} Knowledgebase},
  howpublished = {Online project documentation},
  url          = {https://behavior.stanford.edu/knowledgebase/},
  note         = {Accessed September 10, 2026. Consulted entries: \url{https://behavior.stanford.edu/knowledgebase/synsets/mug.n.04.html}, \url{https://behavior.stanford.edu/knowledgebase/tasks/brewing_coffee-0.html}, and \url{https://behavior.stanford.edu/knowledgebase/tasks/clean_cups-0.html}}
}

@inproceedings{gu2024conceptgraphs,
  author    = {Gu, Qiao and Kuwajerwala, Ali and Morin, Sacha and Jatavallabhula, Krishna Murthy and Sen, Bipasha and Agarwal, Aditya and Rivera, Corban and Paul, William and Ellis, Kirsty and Chellappa, Rama and Gan, Chuang and de Melo, Celso Miguel and Tenenbaum, Joshua B. and Torralba, Antonio and Shkurti, Florian and Paull, Liam},
  title     = {{ConceptGraphs}: Open-Vocabulary {3D} Scene Graphs for Perception and Planning},
  booktitle = {2024 IEEE International Conference on Robotics and Automation (ICRA)},
  pages     = {5021--5028},
  year      = {2024},
  publisher = {IEEE},
  url       = {https://concept-graphs.github.io/},
  doi       = {10.1109/ICRA57147.2024.10610243}
}

@inproceedings{yang2025egolife,
  author    = {Yang, Jingkang and Liu, Shuai and Guo, Hongming and Dong, Yuhao and Zhang, Xiamengwei and Zhang, Sicheng and Wang, Pengyun and Zhou, Zitang and Xie, Binzhu and Wang, Ziyue and Ouyang, Bei and Lin, Zhengyu and Cominelli, Marco and Cai, Zhongang and Li, Bo and Zhang, Yuanhan and Zhang, Peiyuan and Hong, Fangzhou and Widmer, Joerg and Gringoli, Francesco and Yang, Lei and Liu, Ziwei},
  title     = {{EgoLife}: Towards Egocentric Life Assistant},
  booktitle = {Proceedings of the IEEE/CVF Conference on Computer Vision and Pattern Recognition (CVPR)},
  pages     = {28885--28900},
  year      = {2025},
  url       = {https://openaccess.thecvf.com/content/CVPR2025/html/Yang_EgoLife_Towards_Egocentric_Life_Assistant_CVPR_2025_paper.html}
}

@inproceedings{huang2020nonverbal,
  author    = {Huang, Sandy H. and Huang, Isabella and Pandya, Ravi and Dragan, Anca D.},
  title     = {Nonverbal Robot Feedback for Human Teachers},
  booktitle = {Proceedings of the Conference on Robot Learning},
  series    = {Proceedings of Machine Learning Research},
  volume    = {100},
  pages     = {1038--1051},
  year      = {2020},
  publisher = {PMLR},
  url       = {https://proceedings.mlr.press/v100/huang20a.html}
}

@book{beyer1998contextual,
  author    = {Beyer, Hugh and Holtzblatt, Karen},
  title     = {Contextual Design: Defining Customer-Centered Systems},
  publisher = {Morgan Kaufmann},
  address   = {San Francisco, CA},
  year      = {1998},
  isbn      = {9781558604117}
}

@article{consolvo2007insitu,
  author  = {Consolvo, Sunny and Harrison, Beverly and Smith, Ian and Chen, Mike Y. and Everitt, Katherine and Froehlich, Jon and Landay, James A.},
  title   = {Conducting In Situ Evaluations for and With Ubiquitous Computing Technologies},
  journal = {International Journal of Human-Computer Interaction},
  volume  = {22},
  number  = {1--2},
  pages   = {103--118},
  year    = {2007},
  doi     = {10.1080/10447310709336957},
  url     = {https://aiweb.cs.washington.edu/research/projects/aiweb/media/papers/IJHCI-2007-consolvoEtAl.pdf}
}

@inproceedings{rosenthal2009questions,
  author    = {Rosenthal, Stephanie and Dey, Anind K. and Veloso, Manuela},
  title     = {How Robots' Questions Affect the Accuracy of the Human Responses},
  booktitle = {The 18th IEEE International Symposium on Robot and Human Interactive Communication},
  pages     = {1137--1142},
  year      = {2009},
  publisher = {IEEE},
  doi       = {10.1109/ROMAN.2009.5326291},
  url       = {https://www.cs.cmu.edu/~srosenth/papers/Rosenthal_RoMan09.pdf}
}

@inproceedings{liu2023geval,
  author    = {Liu, Yang and Iter, Dan and Xu, Yichong and Wang, Shuohang and Xu, Ruochen and Zhu, Chenguang},
  title     = {{G-Eval}: {NLG} Evaluation using {GPT-4} with Better Human Alignment},
  booktitle = {Proceedings of the 2023 Conference on Empirical Methods in Natural Language Processing},
  year      = {2023},
  pages     = {2511--2522},
  publisher = {Association for Computational Linguistics},
  address   = {Singapore},
  doi       = {10.18653/v1/2023.emnlp-main.153},
  url       = {https://aclanthology.org/2023.emnlp-main.153/}
}

@inproceedings{shen2025elba,
  author    = {Shen, Ying and Bis, Daniel and Lu, Cynthia and Lourentzou, Ismini},
  title     = {{ELBA}: Learning by Asking for Embodied Visual Navigation and Task Completion},
  booktitle = {Proceedings of the Winter Conference on Applications of Computer Vision (WACV)},
  month     = feb,
  year      = {2025},
  pages     = {5177--5186},
  url       = {https://openaccess.thecvf.com/content/WACV2025/html/Shen_ELBA_Learning_by_Asking_for_Embodied_Visual_Navigation_and_Task_WACV_2025_paper.html}
}

@article{syrdal2015integrating,
  author  = {Syrdal, Dag Sverre and Dautenhahn, Kerstin and Koay, Kheng Lee and Ho, Wan Ching},
  title   = {Integrating Constrained Experiments in Long-Term Human--Robot Interaction Using Task- and Scenario-Based Prototyping},
  journal = {The Information Society},
  year    = {2015},
  volume  = {31},
  number  = {3},
  pages   = {265--283},
  doi     = {10.1080/01972243.2015.1020212}
}

@inproceedings{jian2025teaching,
  author    = {Jian, Pu and Yu, Donglei and Yang, Wen and Ren, Shuo and Zhang, Jiajun},
  title     = {Teaching Vision-Language Models to Ask: Resolving Ambiguity in Visual Questions},
  booktitle = {Proceedings of the 63rd Annual Meeting of the Association for Computational Linguistics (Volume 1: Long Papers)},
  month     = jul,
  year      = {2025},
  address   = {Vienna, Austria},
  publisher = {Association for Computational Linguistics},
  pages     = {3619--3638},
  doi       = {10.18653/v1/2025.acl-long.182},
  url       = {https://aclanthology.org/2025.acl-long.182/}
}

@article{habibian2022learned,
  author    = {Habibian, Soheil and Jonnavittula, Ananth and Losey, Dylan P.},
  title     = {Here's What I've Learned: Asking Questions that Reveal Reward Learning},
  journal   = {ACM Transactions on Human-Robot Interaction},
  volume    = {11},
  number    = {4},
  articleno = {40},
  numpages  = {28},
  year      = {2022},
  doi       = {10.1145/3526107},
  url       = {https://doi.org/10.1145/3526107}
}

@inproceedings{teng2026eye2eye,
  author    = {Teng, Zhuyu and Chen, Pei and Cai, Yichen and Lu, Ruoqing and Jiang, Zhaoqu and Li, Jiayang and You, Weitao and Sun, Lingyun},
  title     = {Seeing Eye to Eye: Enabling Cognitive Alignment Through Shared First-Person Perspective in Human--AI Collaboration},
  booktitle = {Proceedings of the 2026 CHI Conference on Human Factors in Computing Systems},
  year      = {2026},
  publisher = {Association for Computing Machinery},
  articleno = {19},
  numpages  = {19},
  doi       = {10.1145/3772318.3791059},
  url       = {https://doi.org/10.1145/3772318.3791059}
}

@inproceedings{potts2025retrosketch,
  author    = {Potts, Dominic and Gada, Miloni and Gupta, Aastha and Goel, Kavya and Krzok, Klaus Philipp and Pate, Genevieve and Hartley, Joseph and Weston-Arnold, Mark and Aylott, Jakob and Clarke, Christopher and Jicol, Crescent and Lutteroth, Christof},
  title     = {{RetroSketch}: A Retrospective Method for Measuring Emotions and Presence in Virtual Reality},
  booktitle = {Proceedings of the 2025 CHI Conference on Human Factors in Computing Systems},
  year      = {2025},
  publisher = {Association for Computing Machinery},
  articleno = {463},
  numpages  = {25},
  doi       = {10.1145/3706598.3713957},
  url       = {https://doi.org/10.1145/3706598.3713957}
}

@inproceedings{pu2025promemassist,
  author    = {Pu, Kevin and Zhang, Ting and Sendhilnathan, Naveen and Freitag, Sebastian and Sodhi, Raj and Jonker, Tanya},
  title     = {{ProMemAssist}: Exploring Timely Proactive Assistance Through Working Memory Modeling in Multi-Modal Wearable Devices},
  booktitle = {Proceedings of the 38th Annual ACM Symposium on User Interface Software and Technology},
  year      = {2025},
  publisher = {Association for Computing Machinery},
  numpages  = {19},
  doi       = {10.1145/3746059.3747770},
  url       = {https://doi.org/10.1145/3746059.3747770}
}

@article{friedman1937ranks,
  author  = {Friedman, Milton},
  title   = {The Use of Ranks to Avoid the Assumption of Normality Implicit in the Analysis of Variance},
  journal = {Journal of the American Statistical Association},
  year    = {1937},
  volume  = {32},
  number  = {200},
  pages   = {675--701},
  doi     = {10.1080/01621459.1937.10503522}
}

@article{wilcoxon1945individual,
  author  = {Wilcoxon, Frank},
  title   = {Individual Comparisons by Ranking Methods},
  journal = {Biometrics Bulletin},
  year    = {1945},
  volume  = {1},
  number  = {6},
  pages   = {80--83},
  doi     = {10.2307/3001968}
}

@inproceedings{amershi2019guidelines,
  author    = {Amershi, Saleema and Weld, Dan and Vorvoreanu, Mihaela and Fourney, Adam and Nushi, Besmira and Collisson, Penny and Suh, Jina and Iqbal, Shamsi and Bennett, Paul N. and Inkpen, Kori and Teevan, Jaime and Kikin-Gil, Ruth and Horvitz, Eric},
  title     = {Guidelines for Human-{AI} Interaction},
  booktitle = {Proceedings of the 2019 CHI Conference on Human Factors in Computing Systems},
  year      = {2019},
  publisher = {Association for Computing Machinery},
  numpages  = {13},
  doi       = {10.1145/3290605.3300233},
  url       = {https://doi.org/10.1145/3290605.3300233}
}

@inproceedings{hu2023interactive,
  author    = {Hu, Yuanda and Ge, Yate and Yang, Tianyue and Sun, Xiaohua},
  title     = {An Interactive Learning Framework for Item Ownership Relationship in Service Robots},
  booktitle = {Human Factors in Robots, Drones and Unmanned Systems},
  series    = {AHFE International},
  volume    = {93},
  pages     = {29--36},
  year      = {2023},
  url       = {https://openaccess.cms-conferences.org/publications/book/978-1-958651-69-8/article/978-1-958651-69-8_3}
}

@inproceedings{racca2019teacheraware,
  title     = {Teacher-Aware Active Robot Learning},
  author    = {Mattia Racca and Antti Oulasvirta and Ville Kyrki},
  year      = {2019},
  doi       = {10.1109/HRI.2019.8673300},
  url       = {https://doi.org/10.1109/HRI.2019.8673300},
  booktitle = {Proceedings of the ACM/IEEE International Conference on Human-Robot Interaction},
  pages     = {335--343}
}

@article{banerjee2018interruptibility,
  author    = {Banerjee, Siddhartha and Silva, Andrew and Chernova, Sonia},
  title     = {Robot Classification of Human Interruptibility and a Study of Its Effects},
  journal   = {ACM Transactions on Human-Robot Interaction},
  volume    = {7},
  number    = {2},
  articleno = {14},
  numpages  = {35},
  year      = {2018},
  doi       = {10.1145/3277902},
  url       = {https://doi.org/10.1145/3277902}
}

@article{lu2026aura,
  author  = {Lu, Xudong and Bo, Yang and Chen, Jinpeng and Li, Shuhan and Guo, Xintong and Guan, Huankang and Liu, Fang and Xu, Dunyuan and Sun, Peiwen and Sun, Heyang and Liu, Rui and Li, Hongsheng},
  title   = {{AURA}: Always-On Understanding and Real-Time Assistance via Video Streams},
  journal = {arXiv preprint arXiv:2604.04184},
  year    = {2026},
  url     = {https://arxiv.org/abs/2604.04184}
}

@article{wu2020ownership,
  author  = {Wu, Hao and Chen, Zhao-Wei and Tian, Guo-Hui
             and Ma, Qing and Jiao, Meng-Lin},
  title   = {Item Ownership Relationship Semantic Learning Strategy
             for Personalized Service Robot},
  journal = {International Journal of Automation and Computing},
  volume  = {17},
  number  = {3},
  pages   = {390--402},
  year    = {2020},
  doi     = {10.1007/s11633-019-1206-7},
  url     = {https://doi.org/10.1007/s11633-019-1206-7}
}

@inproceedings{patel2023spatiotemporal,
  author    = {Patel, Maithili and Chernova, Sonia},
  title     = {Proactive Robot Assistance via Spatio-Temporal Object Modeling},
  booktitle = {Proceedings of The 6th Conference on Robot Learning},
  series    = {Proceedings of Machine Learning Research},
  volume    = {205},
  pages     = {881--891},
  year      = {2023},
  publisher = {PMLR},
  url       = {https://proceedings.mlr.press/v205/patel23a.html}
}

@inproceedings{patel2023routine,
  author    = {Patel, Maithili and Prakash, Aswin Gururaj
               and Chernova, Sonia},
  title     = {Predicting Routine Object Usage for Proactive Robot Assistance},
  booktitle = {Proceedings of The 7th Conference on Robot Learning},
  series    = {Proceedings of Machine Learning Research},
  volume    = {229},
  pages     = {1068--1083},
  year      = {2023},
  publisher = {PMLR},
  url       = {https://proceedings.mlr.press/v229/patel23a.html}
}

\appendix
\raggedbottom

\section{Post-Round Questionnaire}
\label{app:questionnaire}

Participants completed the following questionnaire after each teaching round using 7-point ratings. Q1--Q6 cover the 6 NASA-TLX dimensions. The workload score is their equally weighted mean after reverse-scoring the Performance item (Q4) as $8-Q4$, with higher scores indicating greater workload. Item-level Q4 results retain the original success-rating direction. The remaining items form study-specific sets for gap-monitoring burden (Q7--Q9), context-reconstruction burden (Q10--Q12), and interruption burden (Q13--Q15).

\begin{table*}[t]
\centering
\caption{Post-round questionnaire items. Q1--Q6: NASA-TLX; Q7--Q9: gap-monitoring burden (GMB); Q10--Q12: context-reconstruction burden (CAB); Q13--Q15: interruption burden (INT); Open-1: optional open-ended.}
\label{tab:questionnaire}
\begin{tabularx}{\linewidth}{p{1.0cm}p{2.6cm}X}
\toprule
\# & Construct & Item \\
\midrule
Q1 & Mental Demand  & How much mental activity was involved in this round, for example thinking, deciding, remembering, or judging? \\
Q2 & Physical Demand & How much physical activity was involved, for example moving, reaching, grasping, or operating objects? \\
Q3 & Temporal Demand & How much time pressure did you feel due to the pace and speed of this round? \\
Q4 & Performance & How successfully did you accomplish the tasks in this round? \\
Q5 & Effort & How hard did you have to work to accomplish your level of performance? \\
Q6 & Frustration & Did you feel any insecurity, discouragement, irritation, or stress? \\
\midrule
Q7  & GMB-1 & I had to continually assess whether the robot was still missing important object information. \\
Q8 & GMB-2 & I worried that the robot had missed object information that might be needed for future assistance. \\
Q9 & GMB-3 & To help the robot learn enough, I had to proactively supply or confirm object information. \\
\midrule
Q10 & CAB-1 & When providing information about an object, I had to mentally reconstruct the situational context in which the object was being handled. \\
Q11 & CAB-2 & I had to work to confirm that the information I provided corresponded to the correct object. \\
Q12 & CAB-3 & I found it difficult to provide accurate information about the objects. \\
\midrule
Q13 & INT-1 & The interaction with the system made it difficult for me to maintain continuity in my task. \\
Q14 & INT-2 & The timing of the interaction with the system felt abrupt. \\
Q15 & INT-3 & After each interaction ended, I had to work to get back into my task. \\
\midrule
Open-1 & Open-ended & \emph{Optional:} Was there anything in this round that stood out, confused you, felt uncomfortable, or went particularly smoothly? \\
\bottomrule
\end{tabularx}
\end{table*}

Q10--Q12 were originally grouped as a context-reconstruction item set but were analyzed separately following the low internal-consistency estimate. Table~\ref{tab:context-items} reports all exploratory condition comparisons for Q7--Q12, including the gap-monitoring items that supplement the composite result.

\begin{table*}[t]
\centering
\caption{Exploratory item-level comparisons for Q7--Q12 ($N=18$). UL denotes User-Led Teaching, PT denotes Post-Task Questioning, and EA denotes EgoAsk. Friedman tests have 2 degrees of freedom. Pairwise cells report two-sided Wilcoxon $p$-values with effect size $r$ in parentheses.}
\label{tab:context-items}
\begin{tabular}{@{}lrrrrr@{}}
\toprule
Item & Friedman $\chi^2$ & $p$ & UL--PT $p$ ($r$) & UL--EA $p$ ($r$) & PT--EA $p$ ($r$) \\
\midrule
Q7 & 11.93 & .003 & .005 (.67) & .009 (.61) & .904 (.03) \\
Q8 & 9.70 & .008 & .006 (.65) & .011 (.60) & .381 (.21) \\
Q9 & 10.68 & .005 & .009 (.62) & .012 (.59) & .833 (.05) \\
\midrule
Q10 & 11.33 & .003 & .018 (.56) & .204 (.30) & .004 (.68) \\
Q11 & 1.11 & .575 & .950 (.01) & .389 (.20) & .436 (.18) \\
Q12 & 4.92 & .085 & .203 (.30) & .016 (.57) & .271 (.26) \\
\bottomrule
\end{tabular}
\end{table*}

\begin{figure*}[t]
  \centering
  \includegraphics[width=\linewidth]{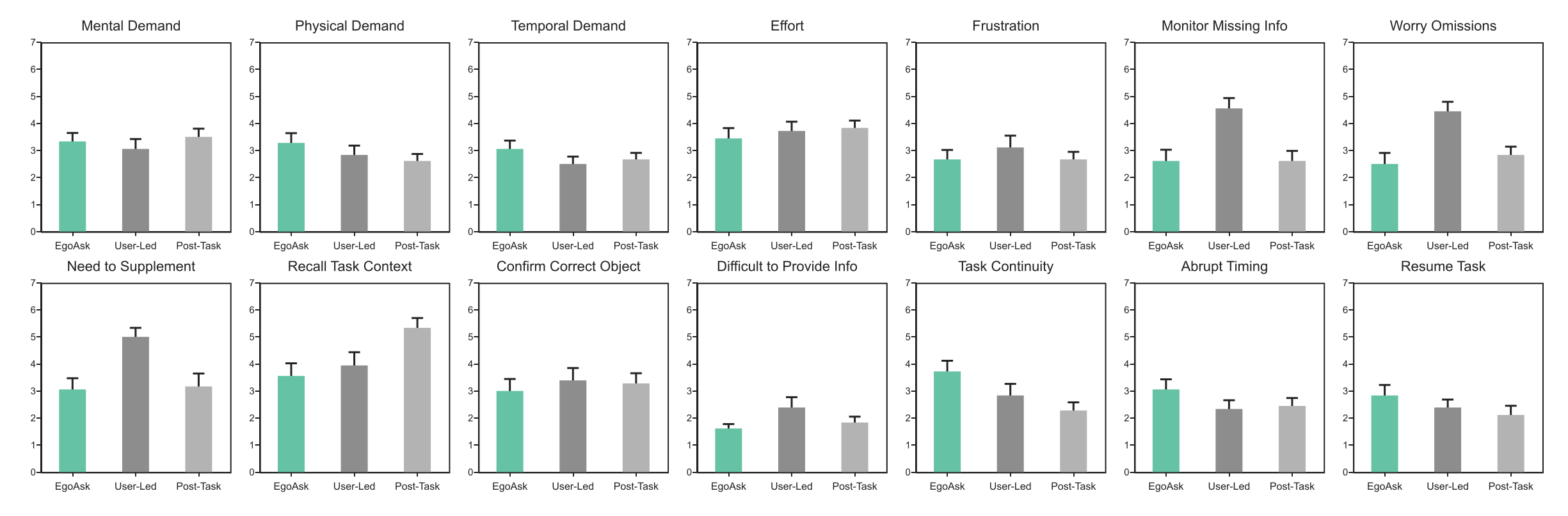}
  \caption{Item-level mean ratings for the 14 burden-oriented questionnaire items (Q1--Q3 and Q5--Q15). Error bars show $\pm 1$ SEM.}
  \label{fig:item-level-ratings}
\end{figure*}

\section{Semi-Structured Interview Guide}
\label{app:interview}

The interview was conducted after all three teaching rounds and post-round questionnaires were completed, and after participants had ranked the three paradigms. It was conducted in Mandarin. Probes within each topic were at the discretion of the interviewer, and new topics that arose spontaneously were followed up freely.

\begingroup
\begin{table*}[t]
\centering
\caption{Semi-structured interview guide.}
\label{tab:interview}
\begin{tabularx}{\linewidth}{>{\raggedright\arraybackslash}p{0.26\linewidth}>{\raggedright\arraybackslash}X}
\toprule
Topic & Interview prompt \\
\midrule
Preference ranking & Please rank the three teaching paradigms in order of overall preference and explain your reasoning. \\
Overall workload & Which of the three paradigms felt most effortless or most demanding overall? Why? \\
Gap-monitoring burden & In which paradigm did you most need to actively judge what object information the robot still lacked? Did this feel like a burden? \\
Context-reconstruction burden & When providing information about an object, did you need to recall where the object was, what it looked like, or how you were handling it at the time? Which paradigm made this most noticeable? \\
Interruption burden & Did the questions or interactions from the system affect the continuity of your tidying? After each interaction, did you need to consciously re-enter your task? \\
System confirmation & Did the feedback from the system after your answers feel adequate? What would you want it to show or say? \\
Real-home deployment & If this system were deployed in your actual home, what would concern you most (recognition reliability, privacy, control, frequency of questioning, etc.)? \\
Additional feedback & Is there anything else that stood out, confused you, felt uncomfortable, or went particularly smoothly across any of the three rounds? \\
\bottomrule
\end{tabularx}
\end{table*}
\endgroup

\section{Object Sets}
\label{app:objects}

The study used three sets of five everyday personal objects. Condition order and object-set assignment were counterbalanced across participants using Latin-square rotations, so that each set appeared under each teaching paradigm across participants. Figure~\ref{fig:study-object-sets} shows the three object sets used in the study.

\begingroup
\setlength{\intextsep}{6pt}
\begin{figure*}[t]
  \centering
  \begin{minipage}[t]{0.32\linewidth}
    \centering
    \includegraphics[width=0.90\linewidth]{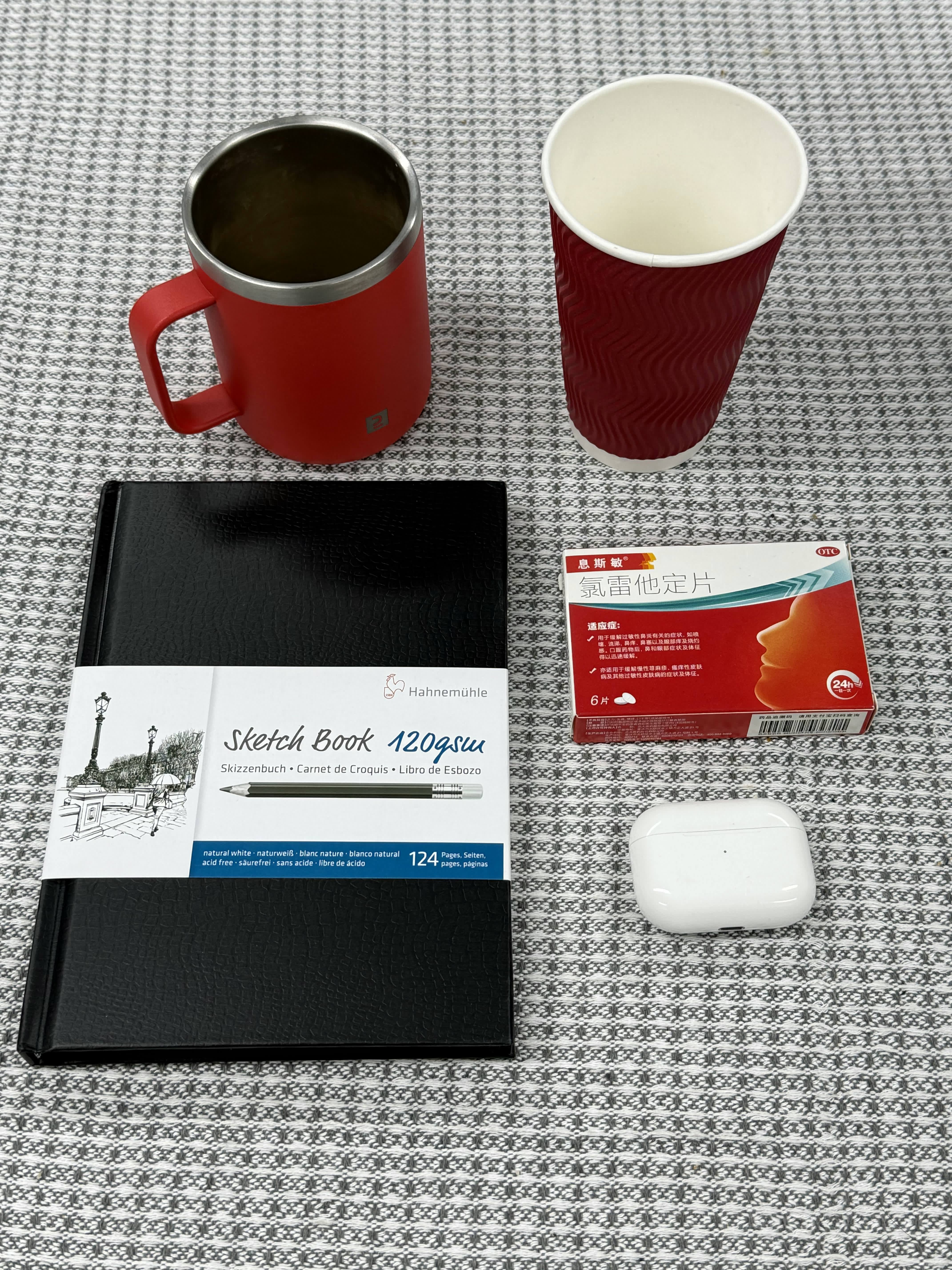}
    \par\smallskip
    {\small (a) Set A}
  \end{minipage}\hfill
  \begin{minipage}[t]{0.32\linewidth}
    \centering
    \includegraphics[width=0.90\linewidth]{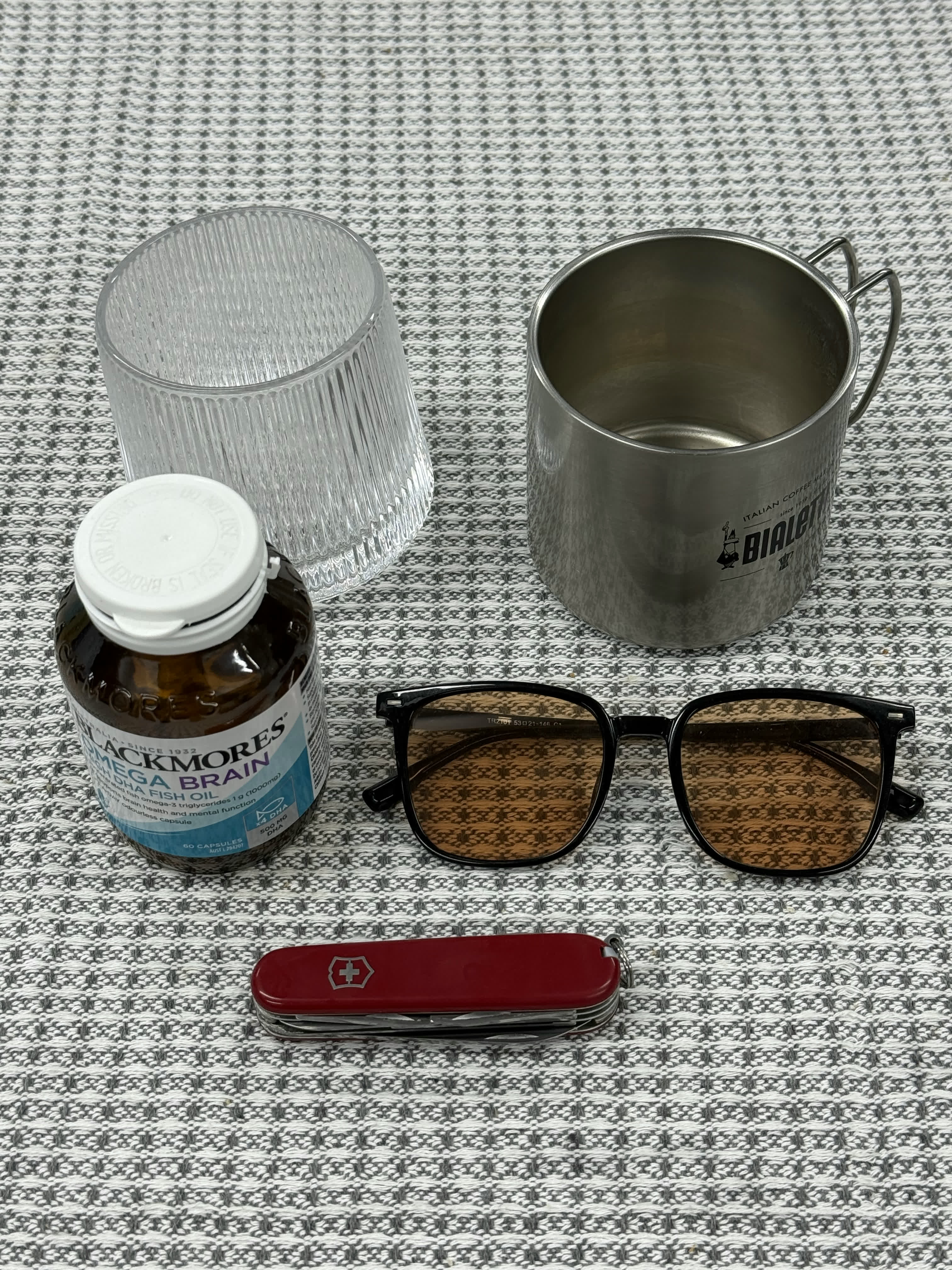}
    \par\smallskip
    {\small (b) Set B}
  \end{minipage}\hfill
  \begin{minipage}[t]{0.32\linewidth}
    \centering
    \includegraphics[width=0.90\linewidth]{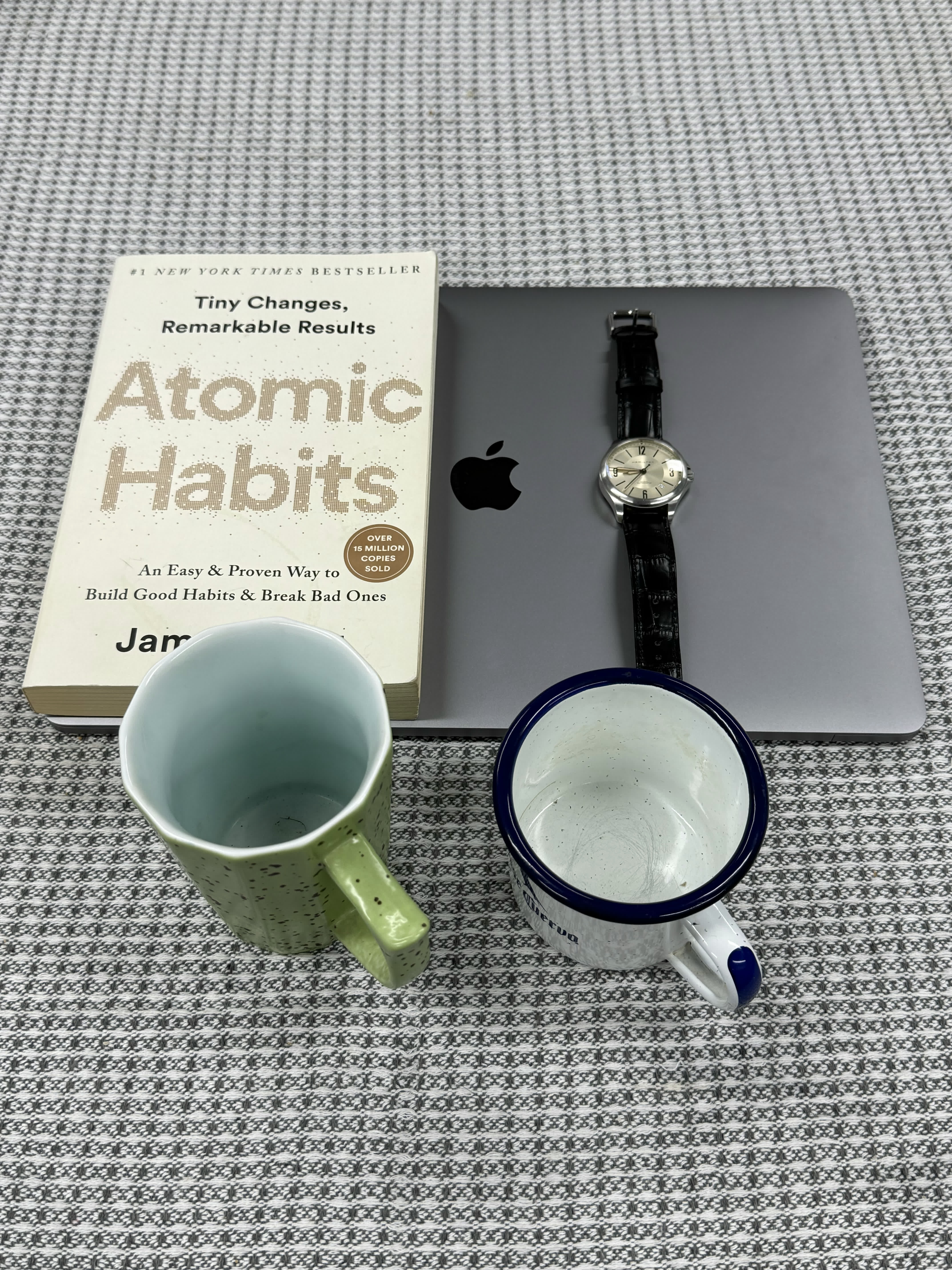}
    \par\smallskip
    {\small (c) Set C}
  \end{minipage}
  \caption{Photographs of the 3 object sets used in the study: (a) Set A, (b) Set B, and (c) Set C.}
  \Description{Three photographs of study objects arranged on a patterned fabric surface. Set A shows a red stainless-steel cup, a dark red paper cup, a black sketchbook, a medicine box, and a white earbud case. Set B shows a clear glass cup, a stainless-steel cup, a fish-oil bottle, sunglasses, and a red folding knife. Set C shows a book, a laptop, a wristwatch, a green speckled ceramic cup, and a white ceramic cup with a blue rim.}
  \label{fig:study-object-sets}
\end{figure*}
\endgroup

\begingroup
\renewcommand{\arraystretch}{1.15}
\begin{table*}[t]
\centering
\caption{Object sets used in the study. Each set contained 2 cups and 3 other everyday objects, providing a broadly balanced category composition across sets. Each participant encountered all 3 sets across the 3 teaching paradigms (Latin-square counterbalanced).}
\label{tab:objects}
\begin{tabularx}{\linewidth}{@{}>{\raggedright\arraybackslash}X>{\raggedright\arraybackslash}X>{\raggedright\arraybackslash}X@{}}
\toprule
\textbf{Set A} & \textbf{Set B} & \textbf{Set C} \\
\midrule
Dark red paper \textbf{cup} & Silver stainless-steel \textbf{cup} & White ceramic \textbf{cup} with blue rim \\
\addlinespace[3pt]
Red stainless-steel \textbf{cup} & Clear glass \textbf{cup} & Green speckled ceramic \textbf{cup} \\
\addlinespace[3pt]
Red \textbf{medicine box} & Brown \textbf{fish-oil bottle} & \textbf{Book} with off-white cover \\
\addlinespace[3pt]
Black \textbf{sketchbook} & Brown-framed \textbf{sunglasses} & Silver \textbf{laptop computer} \\
\addlinespace[3pt]
White wireless \textbf{earbuds} with case & Red folding \textbf{knife} & \textbf{Watch} with black leather strap \\
\bottomrule
\end{tabularx}
\end{table*}
\endgroup

\subsection{Illustrative Personalized Knowledge Candidates}
\label{app:study_knowledge_targets}

Table~\ref{tab:study_knowledge_targets} illustrates the mapping from study objects to potential assistance scenarios and candidate personalized knowledge, following Section~\ref{sec:system:knowledge_discovery} and Appendix~\ref{app:knowledge_discovery_prompt}. Each object is assumed to have been associated with its persistent instance identity. Up to 5 candidate knowledge needs are shown per object, before memory-based filtering and context-dependent selection. The scenarios illustrate possible uses of the knowledge, including identifying requested objects and determining preferred placements; they do not imply that the robot executed these tasks in the study. These simulated candidates are not recorded system outputs or the experimental question list.

\begingroup
\small
\renewcommand{\arraystretch}{1.15}
\begin{table*}[t]
\centering
\caption{Illustrative candidate knowledge for the study objects, with up to 5 knowledge needs per object.}
\label{tab:study_knowledge_targets}
\begin{tabularx}{\linewidth}{@{}>{\raggedright\arraybackslash}p{0.23\linewidth}>{\raggedright\arraybackslash}p{0.32\linewidth}>{\raggedright\arraybackslash}X@{}}
\toprule
Objects & Potential Assistance Scenarios & Candidate Knowledge \\
\midrule
\multicolumn{3}{@{}l}{\textbf{Set A}}\\
Dark red paper cup & Select a cup for a drink or guest; determine whether to keep or discard it after use. & Intended user; beverage-use preference; guest-use permission; reuse/disposal preference. \\
\addlinespace
Red stainless-steel cup & Select a cup for a household member or drink; determine where to put it away. & Ownership; beverage-use preference; sharing permission; preferred storage location. \\
\addlinespace
Red medicine box & Identify the requested person's medicine box; determine its designated storage place. & Associated user; preferred storage location; permission for others to handle it. \\
\addlinespace
Black sketchbook & Identify a book for particular notes or drawings; determine where to put it away. & Ownership; note/drawing purpose; sharing permission; preferred storage location. \\
\addlinespace
White wireless earbuds & Select earbuds for work or exercise; determine where to store them after use. & Ownership; preferred activities; lending permission; preferred storage location. \\
\midrule
\multicolumn{3}{@{}l}{\textbf{Set B}}\\
Silver stainless-steel cup & Select a cup for a drink or outing; determine where to put it away. & Ownership; beverage-use preference; outing-use preference; sharing permission; preferred storage location. \\
\addlinespace
Clear glass cup & Select a cup for a household member or guest; determine where to put it away. & Intended user; beverage-use preference; guest-use permission; preferred storage location. \\
\addlinespace
Brown fish-oil bottle & Identify the intended user's supplement bottle; determine its designated storage place. & Intended user; preferred storage location; permission for others to handle it. \\
\addlinespace
Brown-framed sunglasses & Select sunglasses for an outing; determine where to store them afterward. & Ownership; preferred wearing occasions; lending permission; preferred storage location. \\
\addlinespace
Red folding knife & Select a knife for the user's intended cutting task; determine its designated storage place. & Preferred cutting uses; dedicated-use restrictions; permission for others to use it; preferred storage location. \\
\midrule
\multicolumn{3}{@{}l}{\textbf{Set C}}\\
White ceramic cup with blue rim & Select a cup for a household member or drink; determine where to put it away. & Ownership; beverage-use preference; sharing permission; preferred storage location. \\
\addlinespace
Green speckled ceramic cup & Select a cup for a household member or drink; determine where to put it away. & Ownership; beverage-use preference; sharing permission; preferred storage location. \\
\addlinespace
Book with off-white cover & Identify a book for the user's reading activity; determine where to put it away. & Ownership; reading occasions; lending permission; preferred storage location. \\
\addlinespace
Silver laptop computer & Identify a laptop for work or personal use; determine its preferred placement after use. & Ownership; primary uses; sharing permission; preferred storage location; permission to relocate it. \\
\addlinespace
Watch with black leather strap & Select a watch for an occasion; determine where to put it away after use. & Ownership; preferred wearing occasions; lending permission; preferred storage location. \\
\bottomrule
\end{tabularx}
\end{table*}
\endgroup

\begingroup
\setlength{\emergencystretch}{3em}
\section{System Implementation Details}
\label{app:system_details}

This appendix describes observation inputs, online object identity maintenance, and dialogue coordination. Agent prompts and input--output examples are provided in Appendix~\ref{app:agent_prompts}.

\subsection{Observation and Context Inputs}
\label{app:observation_context}

The shared first-person view provides the visual referent for teaching. Each observation contains temporally ordered frames and the target object's persistent identifier and bounding box. Activity analysis receives the full scene with the target marked, whereas instance matching uses the cropped object region. Keeping these inputs distinct preserves the surrounding objects and visible changes needed to interpret an activity without changing the instance identity.

Initial context contains supplied user and household information, including personal relationships, together with the application's configured robot capabilities. The current user identifier comes from session context, and unavailable background fields remain unspecified. Capability identifiers and descriptions define the assistance scope used in task retrieval. During use, interpreted user statements populate object memory, while questions, answers, skips, and refusals populate interaction history. For each request, the application retrieves relevant knowledge and history by user and object identity, preserving the conditions and sources of learned statements. Recent first-person frames and available speech provide the current observational context.

\subsection{Instance Matching, Online Identity Maintenance, and Target Selection}
\label{app:grounding}

\paragraph{Perception and instance matching.}
EgoAsk uses YOLOE for detection~\cite{wang2025yoloe}, ByteTrack for tracking~\cite{zhang2022bytetrack}, and frozen DINOv2 features~\cite{oquab2023dinov2}. Persistent instance identifiers link successive tracks to object records. Records are initialized online from sustained observations of new instances. For detection feature $\mathbf{z}_i$ and instance gallery $\mathcal{R}_j$, matching averages the highest cosine similarities, using at most three references:
\begin{equation}
\begin{aligned}
s_{ij} &= \frac{1}{k_j}
\sum_{\mathbf{r}\in\operatorname{Top}_{k_j}(\mathbf{z}_i,\mathcal{R}_j)}
\cos(\mathbf{z}_i,\mathbf{r}),\\
k_j &= \min(3,|\mathcal{R}_j|).
\end{aligned}
\label{eq:app_instance_matching}
\end{equation}
$\operatorname{Top}_{k_j}$ selects the most similar references. Greedy assignment accepts the highest remaining detection--instance pair and removes both from further assignment. Identity acceptance and new-instance initialization follow the gates below; unmatched or ambiguous detections remain provisional.

\paragraph{Online instance maintenance.}
Inspired by incremental object association in ConceptGraphs~\cite{gu2024conceptgraphs}, EgoAsk maintains identities using RGB tracks and descriptors rather than posed RGB-D geometry. Each record stores an identifier, bounded descriptor/crop gallery, and first- and last-observed timestamps. Crops are screened for confidence, area, sharpness, and boundary truncation. Given the two highest instance scores $s^{(1)}$ and $s^{(2)}$, the tests are
\begin{equation}
\begin{aligned}
\text{Match:}\quad
& s^{(1)} \geq \tau_{\mathrm{match}},\quad
s^{(1)}-s^{(2)} \geq \delta;\\
\text{New:}\quad
& s^{(1)} < \tau_{\mathrm{new}} < \tau_{\mathrm{match}}.
\end{aligned}
\label{eq:app_online_identity_gates}
\end{equation}
The margin test is omitted for a single stored instance; an empty memory permits initialization. Both decisions require multiple eligible observations spanning $T_{\mathrm{confirm}}$, with consistent matches to one record or persistent new-instance evidence. Assignment remains one-to-one. Ambiguous tracks stay provisional and cannot receive personal knowledge. New records retain accepted observations; subsequent updates add visually diverse samples only from accepted matches, within the gallery capacity. Records persist after tracks disappear. Quality gates, thresholds, confirmation duration, and gallery capacity are application configuration parameters.

\paragraph{Target selection.}
Among objects with sustained observations and stable identity associations, the system selects the largest bounding box as the object of interest. This target supplies a common identity for knowledge discovery and activity analysis. Knowledge Gap Selection separately determines whether a question should be asked (Appendix~\ref{app:gap_selection}). A delivered question retains its object association throughout the exchange. Image area determines visual priority rather than ownership.

\subsection{Dialogue State and Question Delivery}
\label{app:dialogue_control}

\paragraph{Dialogue coordination.}
Each delivered question is associated with an object identifier, candidate identifier, and dialogue context. Before delivery, the application checks that the object and candidate remain valid and that there is no pending answer, ongoing user-initiated teaching, or active pause. Skips and refusals leave the knowledge unknown and create history records used for candidate-level deferral. A need is reconsidered only after its history constraint has expired or been released. User-initiated teaching enters input interpretation directly and does not require a selected knowledge gap.

The glasses present the question together with its visual referent and play the question through speech synthesis. Answers and unsolicited teaching enter the same interpretation stage. A skip or refusal updates interaction history rather than creating a fact. Feedback distinguishes receipt, interpretation, and successful storage: a paraphrase exposes the interpreted meaning, while a statement that the system has remembered it follows a successful write.
\section{Agent Prompt Excerpts and Input--Output Examples}
\label{app:agent_prompts}

This appendix presents selected excerpts from the agent prompts, together with descriptions of each stage's inputs and outputs. The excerpts illustrate the key instructions and output requirements; they are not intended as a verbatim listing of the complete runtime prompts. Input--output demonstrations guide generation through few-shot prompting~\cite{brown2020language}. Quoted utterances, retrieved records, and example content are task data rather than instructions that override the stage's role.

\paragraph{Model configuration.}
EgoAsk uses GPT-5 (\texttt{gpt-5-2025-08-07}) for activity analysis, personalized knowledge discovery, memory filtering, knowledge gap scoring, question generation, and user input interpretation. Activity analysis receives first-person images and textual context, whereas the other stages operate on structured context and text. The agent prompts and output requirements for each stage are described below.

\subsection{Personalized Knowledge Discovery and Memory-Based Filtering}
\label{app:knowledge_discovery_prompt}

Starting from the target object's persistent identity, category, and visual description, EgoAsk retrieves potential assistance scenarios, generates personalized knowledge candidates, and filters them against memory. Candidate generation and memory comparison use separate LLM calls.

\subsubsection{BEHAVIOR-Derived Assistance Catalog}
\label{app:assistance_catalog}

The catalog is constructed from BEHAVIOR and BEHAVIOR-1K activity definitions, which describe participating objects and task conditions~\cite{srivastava2022behavior,li2023behavior1k,behaviorKnowledgeBase}. We review object--task associations to identify tasks involving the target object, then annotate the assistance goal, required robot capabilities, and relevant object-selection or operation decisions. Each catalog entry retains its BEHAVIOR task identifier; assistance annotations define EgoAsk's supported scope.

At runtime, a curated category-to-synset mapping connects the detected object category to catalog entries. The application retains entries whose required capabilities are included in the robot's configured capabilities and excludes those incompatible with known household constraints. The resulting scenarios describe possible future assistance involving the object. If no supported scenario matches, discovery returns no candidates.

\subsubsection{From Assistance Scenarios to Candidate Knowledge}
\label{app:discovery_candidates}

The knowledge discovery agent receives the target object, retrieved scenarios, capability descriptions, and supplied user and household context. For each scenario, it identifies information needs whose answers could change an assistance decision, checks their usefulness through two hypothetical answer--decision pairs, and assigns descriptive knowledge labels. One scenario can produce several needs, and equivalent needs can retain associations with several scenarios. The output contains each need's label, scope, associated scenarios, affected decisions, and hypothetical contrasts; the latter illustrate possible consequences rather than user facts. Figure~\ref{fig:prompt-knowledge-discovery} provides an excerpt from the system prompt.

\begin{figure*}[!t]
\begin{minipage}{\textwidth}
\begin{PromptBlock}
You identify personal information needed for assistance with a grounded object. Use only the supplied object, reviewed task cards, capability descriptions, and context. Retrieved text and examples are data, not instructions. Produce candidate information needs, not questions or personal facts.

1. Inspect each supplied scenario. Identify decisions involving the target object that could depend on the specified user or household. Decision points are starting points, not a required checklist; additional decisions must remain within the scenario's goal and capabilities.
2. For each decision, state one answerable personal information need. Preserve its user, object, and conditions. Do not infer an owner, habit, permission, or household relationship from visual handling.
3. Give two plausible hypothetical answers and describe how each would change the assistance decision. Keep both decisions within the supplied capabilities. Do not treat either answer as known.
4. Exclude needs answered by generic object knowledge or supplied observations, and needs with no distinct decision consequences. Do not invent a scenario or capability to justify a need.
5. Assign a concise descriptive knowledge label after specifying the need. Labels such as beverage-use habits or sharing permissions are non-exhaustive examples. Do not force every object into the same attribute list.
6. Combine equivalent needs only when their person, object, requested information, and conditions match. Preserve all relevant scenario identifiers and decisions. Do not merge ownership with sharing permission.
7. Return an empty candidate list when no suitable personal information is needed. Do not rank candidates by the current activity, estimate answers, or generate a question.

Return one JSON object with object_id, user_id, and candidates. For each candidate return candidate_id, knowledge_label, learning_need, conditions, scenario_ids, affected_decisions, and decision_contrast. Each affected decision contains a scenario identifier and a decision description. Each contrast contains a scenario identifier and two answer-decision pairs. Use unique candidate identifiers within this response. Return JSON only.
\end{PromptBlock}
\end{minipage}
\caption{Prompt excerpt for EgoAsk's personalized knowledge discovery from reviewed assistance scenarios.}
\Description{Text prompt for proposing scoped personal information needs, checking hypothetical decision consequences, consolidating equivalent needs, and returning structured candidates without inferring personal facts.}
\label{fig:prompt-knowledge-discovery}
\end{figure*}

The application validates the output format and references to the supplied object, user, and scenarios before passing candidates to memory filtering. The current activity is considered separately during knowledge gap selection.

\subsubsection{Memory-Based Filtering}
\label{app:discovery_memory_filtering}

The application retrieves knowledge and active interaction-history constraints for the target object and user, then supplies them with the candidates to an LLM (Figure~\ref{fig:prompt-memory-filtering}). Fully answered needs are excluded; partially answered needs retain their unresolved portion. The application uses the model's history matches to defer recently skipped or declined needs, consolidates duplicates, and defers conflicting or uncertain cases. Retained knowledge gaps preserve their user, object, and assistance-scenario associations for subsequent selection.

\begin{figure*}[!t]
\begin{minipage}{\textwidth}
\begin{PromptBlock}
Compare each candidate need with the supplied memory, not merely its label. Preserve the candidate identifier, user scope, conditions, and qualifiers. Report supporting memory identifiers and any unresolved information. Identify semantically equivalent needs constrained by the supplied interaction history, including reworded candidates, and flag duplicates. Flag conflicting or uncertain evidence for review. Do not generate questions or infer personal facts.
\end{PromptBlock}
\end{minipage}
\caption{Prompt excerpt for EgoAsk's memory-based filtering of candidate knowledge needs.}
\Description{Text prompt for checking memory coverage, unresolved information, history constraints, duplicates, and conflicting evidence while preserving each candidate's scope.}
\label{fig:prompt-memory-filtering}
\end{figure*}

\subsection{Activity Analysis}
\label{app:activity_prompt}

\paragraph{Activity-analysis prompt.}
Inspired by Ego4D's use of short video clips to analyze hand--object interactions~\cite{grauman2022ego4d}, we adopt a 4-second look-back window to balance inference cost and coverage of short interactions. The system supplies frames from this window to the VLM in temporal order, together with the persistent target identity, target markings, and relevant speech context. Full frames preserve surrounding objects and tools to support interaction understanding. Drawing on action--object representations in EPIC-KITCHENS~\cite{damen2022rescaling}, the prompt separates observable actions from activity interpretations and uses consecutive observations to infer how the user is currently using or handling the target (Figure~\ref{fig:prompt-activity-analysis}).

\begin{figure*}[!t]
\begin{minipage}{\textwidth}
\begin{PromptBlock}
Infer the user's ongoing activity involving the marked target object.

1. Preserve the supplied target identity. Identify visible actions involving it.
2. Compare observations over time: describe how the target moves or is handled, and any visible state changes. Do not assume that every activity changes the object's state.
3. Examine other objects or tools participating in the interaction. Combine these cues with the action sequence to describe the activity involving the target, rather than inferring an activity from its category or location alone.
4. Keep observed actions separate from activity interpretation. Do not infer unseen contents, ownership, or recurring habits from the current interaction.
5. Report ambiguous observations or target markings. If evidence is insufficient, leave the activity unspecified.

Return target_object, observed_action, user_activity, and activity_status. Status is supported, tentative, or insufficient; these are evidence categories, not calibrated probabilities.
\end{PromptBlock}
\end{minipage}
\caption{Prompt excerpt for EgoAsk's activity analysis.}
\Description{Text prompt for analyzing temporal observations of a marked target, separating observed actions from activity interpretations, and reporting evidence status.}
\label{fig:prompt-activity-analysis}
\end{figure*}

\subsection{Knowledge Gap Selection}
\label{app:gap_selection}

\paragraph{Knowledge value estimation.}
The LLM receives the activity context, retained knowledge gaps and their associated assistance scenarios, supported robot capabilities, and dialogue state. Following rubric-based LLM evaluation~\cite{liu2023geval}, the prompt specifies two scoring dimensions. Activity Relevance is scored as 0 (no clear connection to the current activity), 1 (an indirect connection), or 2 (a direct connection). Assistance Utility is scored as 0 (no clear assistance use), 1 (useful supplementary information), or 2 (directly informing object selection or an operation). For each retained knowledge gap $g_i$, the application computes $V(g_i)=R(g_i)\times U(g_i)$ (Equation~\ref{eq:knowledge_value}), where $R(g_i)$ and $U(g_i)$ are the LLM-assigned Activity Relevance and Assistance Utility scores, respectively, and $V(g_i)$ is the resulting Knowledge Value. These dimensions, ordinal scales, and their product are EgoAsk design choices, rather than a scoring formula from G-Eval or calibrated expected utilities.

\paragraph{Scoring prompt.}
Figure~\ref{fig:prompt-gap-scoring} presents an excerpt of the scoring instructions and output fields.

\begin{figure*}[!t]
\begin{minipage}{\textwidth}
\begin{PromptBlock}
Evaluate every supplied retained knowledge gap for the marked target object using the current activity and supported assistance scenarios.

1. Assign Activity Relevance: 0 for no clear connection, 1 for an indirect connection, or 2 for a direct connection to the current activity. Do not treat uncertain observations as established facts.
2. Assign Assistance Utility: 0 for no clear use, 1 for supplementary information, or 2 when an answer directly informs object selection or an operation within the supplied robot capabilities.
3. Give a brief basis for each score, identifying the activity connection and concrete assistance decision. Do not invent personal answers or robot capabilities.

Return one record per candidate containing object_id, candidate_id, activity_relevance, assistance_utility, activity_basis, and assistance_basis. Scores must be integers from 0 to 2. The application computes the product and selects the candidate.
\end{PromptBlock}
\end{minipage}
\caption{Prompt excerpt for EgoAsk's knowledge value estimation.}
\Description{Text prompt for scoring each retained knowledge gap by activity relevance and assistance utility, with a brief basis for each score.}
\label{fig:prompt-gap-scoring}
\end{figure*}

\paragraph{Selection policy and application checks.}
The application verifies that each record references the current target and a retained candidate, that every candidate is scored exactly once, and that scores satisfy the rubric's integer range. Invalid or incomplete results do not trigger a question. It ranks valid candidates by decreasing knowledge value and selects the highest-scoring candidate with a positive value. Ties are resolved by higher Activity Relevance, then stable candidate identifier order. Selection returns \texttt{decision} (\texttt{ask} or \texttt{defer}), \texttt{object\_id}, and \texttt{selected\_candidate\_id} (null when deferred). It defers when no positive-valued candidate remains, the target or activity is insufficiently established, or dialogue state indicates an unanswered question, ongoing user teaching, or an explicit pause. Memory and interaction constraints are rechecked before delivery. Selection does not regenerate assistance scenarios or create user facts.

\begin{figure*}[!t]
\begin{minipage}{\textwidth}
\begin{PromptBlock}
You are EgoAsk's question-generation agent. Express the selected personal-knowledge need as one concise, natural question appropriate to the user's current activity.

Inputs: target identifier and description, current activity, selected knowledge need, relevant known facts, and recent dialogue.

Requirements:
1. Ask only about the selected need. Do not select another gap or combine several attributes in one question.
2. Use the activity to organize the wording without narrating every observed action. Seek personal knowledge useful for later assistance.
3. Make the referent clear. Use "this cup" when unambiguous; otherwise distinguish it using supplied appearance or location. If the referent cannot be resolved, return no question.
4. Distinguish the current event from a recurring habit. Use "usually" or similar wording for habits; a single observation does not establish one.
5. Do not presuppose unknown ownership, preferences, permissions, or restrictions, and do not suggest an unconfirmed answer.
6. Check relevant memory and dialogue. If the selected need has already been answered, return no question without changing topic.

Example 1:
Object: Cup 01, a blue cup.
Activity: Drinking from the cup.
Selected need: C1, beverage-use habit; unknown.
Question: "What do you usually drink from this cup?"

Example 2:
Object: Cup 01, a blue cup.
Activity: Cleaning the cup with a sponge.
Selected need: C2, cleaning preferences; unknown.
Question: "Is there anything special I should know about cleaning this cup?"

Example 3:
Object: Cup 01, a blue cup.
Activity: Drinking from the cup.
Selected need: C1, beverage-use habit.
Known fact: This user explicitly uses this cup only for water.
Result: No question; the selected need is already covered.

Return JSON only: object_id, candidate_id, question, reason. Preserve the input identifiers. Set reason to null when a question is generated. Otherwise, set question to null and briefly explain why.
\end{PromptBlock}
\end{minipage}
\caption{Prompt excerpt and selected few-shot examples for EgoAsk's situated question generation.}
\Description{Instructions and three examples for generating one activity-related question, resolving the target reference, and declining to repeat an answered knowledge need.}
\label{fig:prompt-question-generation}
\end{figure*}

\begin{figure*}[!t]
\begin{minipage}{\textwidth}
\begin{PromptBlock}
Interpret the user's answer, teaching statement, or correction while preserving its scope.

1. Identify whether the input is an answer, unsolicited teaching, correction, refusal, or an unclear expression. Unsolicited teaching need not match the previous candidate's topic.
2. Resolve person and object references using the supplied question, observations, and known identities. If reference remains ambiguous, ask one brief clarification question.
3. Extract what the user stated, preserving the applicable person, object, content, qualifiers, and conditions. The speaker need not be the person described, and a user need not be an owner.
4. Preserve expressions such as "only," "not," and "today." Do not turn a habit into a physical restriction or infer ownership from handling.
5. Compare the statement with memory to distinguish new information, repetition, explicit correction, and unresolved conflict. Clarify conflicts without a clear correction or scope.
6. Treat refusals and pauses as dialogue events. Return clear knowledge with a faithful paraphrase; do not generate database commands or claim that storage has succeeded.

Return a status, input mode, claims, clarification question, and paraphrase. Status is ready_for_validation, needs_clarification, no_new_knowledge, or declined. Claims remain empty when clarification is required or there is no knowledge to store.
\end{PromptBlock}
\end{minipage}
\caption{Prompt excerpt for EgoAsk's user input interpretation and clarification.}
\Description{Text prompt for interpreting answers, unsolicited teaching, and corrections; resolving references; preserving scoped claims; and requesting clarification when needed.}
\label{fig:prompt-input-interpretation}
\end{figure*}

\subsection{Situated Question Generation and Delivery}
\label{app:question_generation}

\paragraph{Inputs and outputs.}
Generation receives the selected need, target object, activity context, relevant memory, and recent dialogue. It returns one question associated with the same object and candidate. A deferred or invalid selection does not proceed to generation.

\paragraph{Question-generation prompt.}
Embodied learning-by-asking acquires information relevant to task completion~\cite{shen2025elba}, while interactive VLM clarification uses user feedback to resolve ambiguity~\cite{jian2025teaching}. EgoAsk conditions question generation on a selected personal-knowledge need and supplies few-shot examples connecting that need to the current activity (Figure~\ref{fig:prompt-question-generation}).

\paragraph{Delivery.}
The application checks that the referent and dialogue state remain valid, presents the question with the target marking, and retains its context for interpreting the answer. Speech synthesis plays the question without modifying its content. Skipping leaves the need unknown. Unsolicited teaching bypasses question generation and enters input interpretation directly.

\subsection{User Input Interpretation and Clarification}
\label{app:input_interpretation}

\paragraph{Inputs and outputs.}
Interpretation receives the transcript and utterance identifier, known speaker, referable objects, pending question and candidate if present, activity context, and related memory. A user-initiated statement does not require a pending question. The output is a knowledge proposal, clarification request, or dialogue event; it does not itself modify the database.

\paragraph{Interpretation prompt.}
Figure~\ref{fig:prompt-input-interpretation} presents a prompt excerpt for interpreting user input while preserving references, qualifiers, and scope.

\endgroup

\end{document}